\documentclass[12pt]{article}
\usepackage{mathbbol}

\def\beq{\begin{equation}}
\def\eeq{\end{equation}}
\def\bea{\begin{array}}
\def\eea{\end{array}}
\def\beqa{\begin{eqnarray}}
\def\eeqa{\end{eqnarray}}
\def\myIm{{\Im m}}

\def\Pexp{{\rm Pexp}}
\relax
\relax
\relax

\def\cO{{\cal{O}}}
\def\cP{{\cal{P}}}
\def\cA{{\cal{A}}}
\def\cF{{\cal{F}}}
\def\cG{{\cal{G}}}
\def\ddz{{\frac{d}{dz}}}
\def\Tr{{\rm Tr}}
\def\tr{{\rm tr}}
\def\diag{{\rm diag}}
\def\pl{{{\cal P}_\infty}}

\def\SD{self\--du\-al}
\def\ASD{an\-ti\--\-self\-du\-al}
\newcommand{\re}{\relax{\rm I\kern-.18em R}}

\newcommand{\refeq}[1]{\mbox{Eq.~(\ref{eq:#1})}}
\newcommand{\half}{{\scriptstyle{{1\over 2}}}}
\newcommand{\hhalf}{{\scriptstyle{{1/2}}}}
\newcommand{\quarter}{{\scriptstyle{{1\over 4}}}}
\newcommand{\veps}{\varepsilon}
\newcommand{\zahlen}{{\mathbb{Z}}}
\newcommand{\ein}{{\Eins}}

\begin{document}

\hfill\hbox{INLO-PUB-05/02}
\vskip5mm
\begin{center}{\Large\bf Multi-Caloron solutions}\\[1cm]
{\bf Falk Bruckmann} and {\bf Pierre van Baal}\\[3mm]
{\em Instituut-Lorentz for Theoretical Physics, University
of Leiden,\\
P.O.Box 9506, NL-2300 RA Leiden, The Netherlands}
\end{center}

\section*{Abstract}
We discuss the construction of multi-caloron solutions with non-trivial
holonomy, both as approximate superpositions and exact self-dual solutions.
The charge $k$ $SU(n)$ moduli space can be described by $k n$ constituent 
monopoles. Exact solutions help us to understand how these constituents can 
be seen as independent objects, which seems not possible with the approximate 
superposition. An ``impurity scattering" calculation provides relatively 
simple expressions. Like at zero temperature an explicit parametrization 
requires solving a quadratic ADHM constraint, achieved here for a class of 
axially symmetric solutions. We will discuss the properties of these exact 
solutions in detail, but also demonstrate that interesting results can be 
obtained without explicitly solving for the constraint. 

\section{Introduction}

The last four years more understanding has been gained of the interplay 
between instantons and monopoles in non-Abelian gauge theories, based 
on the ability to construct exact caloron solutions, i.e. instantons at 
finite temperature for which $A_0$ approaches a constant at spatial
infinity~\cite{KvB2,KvBn}. This last condition is best expressed by 
specifying the Polyakov loop to approach a constant value at infinity, 
also called the holonomy. It is the finite action that demands the 
field strength to go to zero at infinity, and guarantees the Polyakov 
loop to be independent of the direction in which we approach infinity. 
One can parametrize this Polyakov loop in terms of its eigenvalues 
$\exp(2\pi i\mu_j)$ and a gauge rotation $g$, such that
\beq
\pl=\lim_{x\to\infty}\Pexp(\int_0^\beta A_0(\vec x,t)dt)=g^\dagger
\exp(2\pi i\diag(\mu_1,\mu_2,\ldots,\mu_n))g,
\eeq
which can be arranged such that $\sum_{i=1}^n\mu_i=0$, and $\mu_1\leq\mu_2
\leq\ldots\leq\mu_n\leq\mu_{n+1}$, with $\mu_{n+k}\equiv 1+\mu_k$. This 
is in the gauge where the gauge fields, assumed to be anti-hermitian matrices
taking values in the algebra of $SU(n)$, are periodic in the time direction 
$A_\mu(\vec x,t)=A_\mu(\vec x,t+\beta)$. In our conventions the field 
strength is given by $F_{\mu\nu}(x)=\partial_\mu A_\nu(x)-\partial_\nu 
A_\mu(x)+[A_\mu(x),A_\nu(x)]$.

We will find it convenient to construct the multi-caloron solutions from 
the ADHM-Nahm Fourier construction~\cite{KvB2,KvBn,ADHM,NCal} based 
on taking instantons in $R^4$, which periodically repeat in the 
time direction (see also Ref.~\cite{Lee} which uses directly the Nahm 
transformation~\cite{NCal,Nahm} as the starting point for the charge 1 
construction). To allow for non-trivial holonomy, the periodicity is only 
up to a constant gauge rotation (which is the holonomy). In this so-called 
algebraic gauge, all gauge field components vanish at spatial infinity, and 
we may approximately superpose these calorons by simply adding the gauge 
fields. When each gauge field is periodic up to the same constant gauge 
transformation, the sum satisfies the same property. It should be noted that 
we are not allowed to add gauge fields with different holonomy; in an {\em 
infinite} volume the holonomy is fixed by the boundary condition. As to the 
topological charge, we recall that in the Atiyah-Drinfeld-Hitchin-Manin 
(ADHM) construction~\cite{ADHM} it is supported by gauge singularities (the 
algebraic gauge is for this reason also called the singular gauge), as 
opposed to at infinity being a pure gauge with the gauge function having the 
appropriate winding number. To deal with the gauge singularity of one 
instanton, when adding the field of the others, one has to smoothly 
deform the gauge field of the latter to vanish near the gauge singularity.
As long as the singularities are not too close, this can be done without 
a significant increase in the action. On the lattice this problem does not 
occur, when hiding singularities between the meshes of the lattice.

\begin{figure}[htb]
\vskip4.1cm
\includegraphics{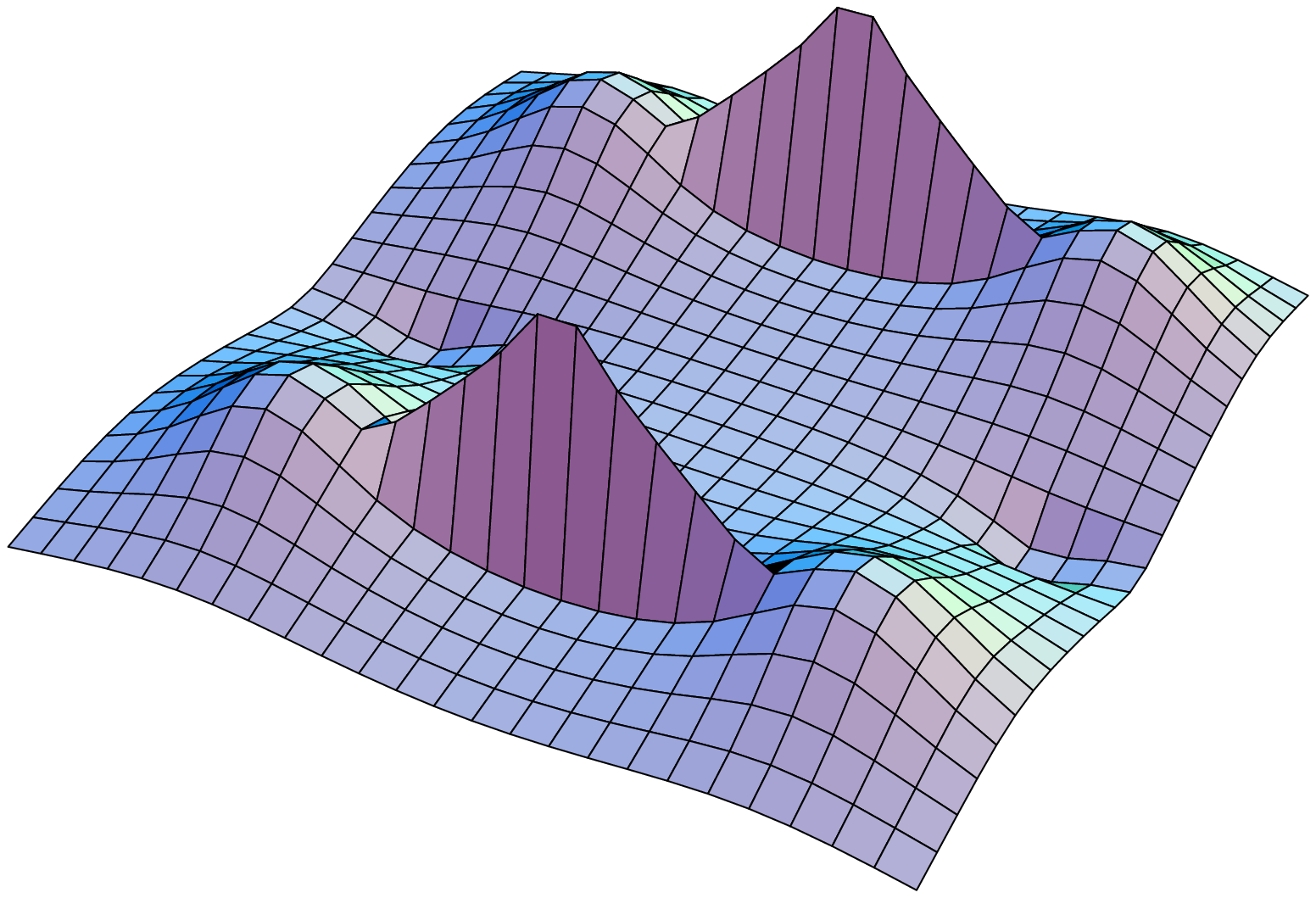}
\includegraphics{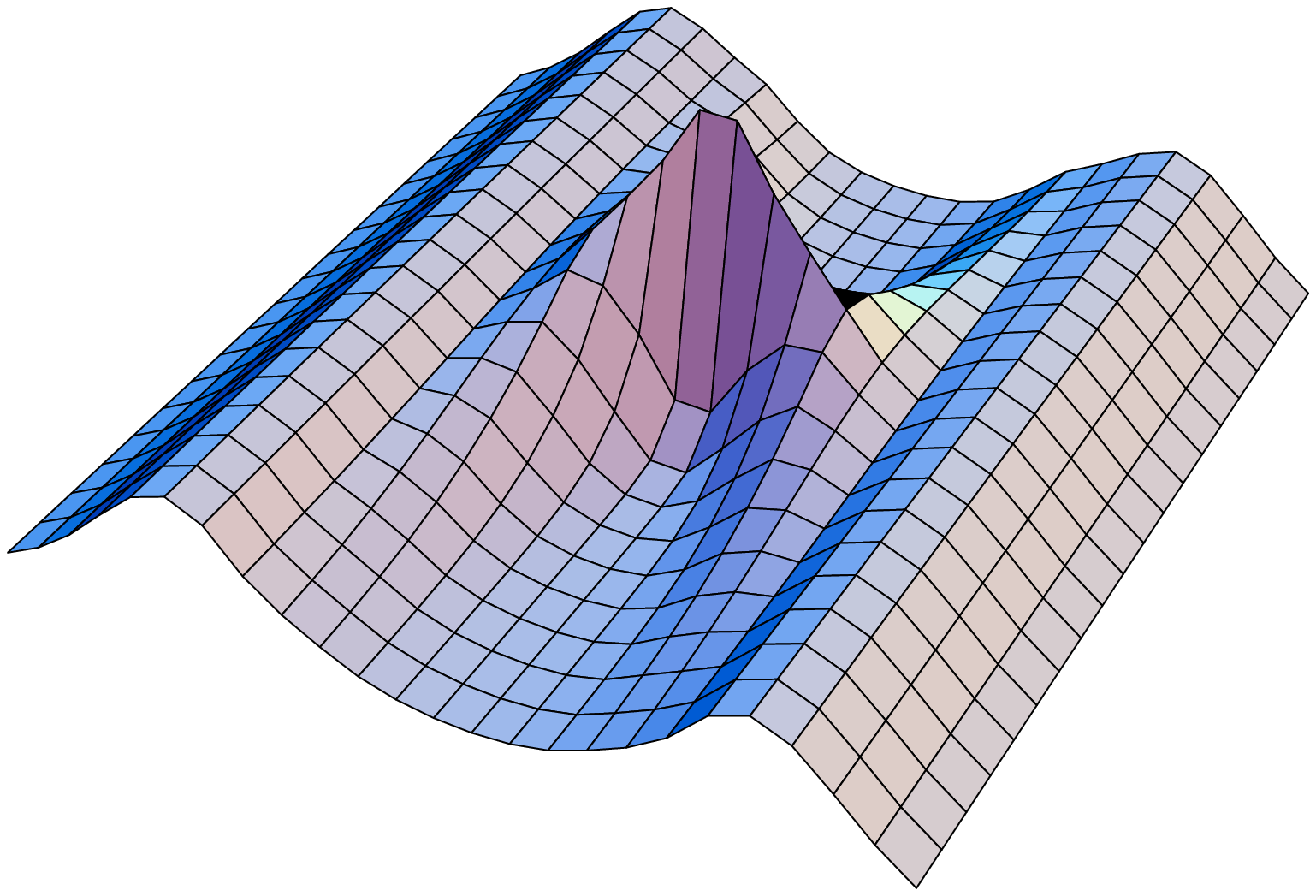}
\caption{Approximate superposition of two $SU(2)$ charge 1 calorons with its 
pairs of equal mass constituents at $\vec x=(2,0,2),~(2,0,8)$ and $\vec x=(8,
0,2),~(8,0,8)$. The logarithm of the action density is plotted as a function 
of $x$ and $z$. The plot on the right shows one of the would-be Dirac strings, 
zooming in by a factor 40 on the transverse direction.}\label{fig:String}
\end{figure}

Calorons, however, have an additional feature. When squeezed in the imaginary 
time direction (the size $\rho$ becoming bigger than the period $\beta$), 
they split in constituent monopoles with masses $8\pi^2\nu_j/\beta$ ($\nu_j
\equiv\mu_{j+1}-\mu_j$). Outside the cores of these monopoles the gauge field 
becomes abelian. Ignoring the charged components, which decay exponentially 
outside the cores of the constituent monopoles, the field is described in 
terms of self-dual Dirac monopoles. The singularity of a Dirac string can 
only be avoided by {\em not} neglecting the contributions coming from the 
charged components, even when far away from the constituents. For a single 
caloron we would not care, since the Dirac string is not seen in gauge 
invariant quantities. It does, however, involve a rather subtle interplay 
between the charged and neutral components of the gauge field in the vicinity 
of the would-be Dirac string~\cite{KvB2}. It is this subtle interplay that is 
disturbed when we add gauge fields of various calorons together. Unlike for 
the gauge singularity, the combined field will not diverge, but it shows a 
narrow and steep enhancement at the location of the would-be Dirac string as 
illustrated in Fig.~\ref{fig:String}, where we added two $SU(2)$ calorons. 
Also here one may shield the Dirac strings from these tails. But one always 
pays the price that the Dirac string no longer can be hidden, and carries 
energy. Let us stress again that these are genuine gauge invariant non-singular
features in the configuration, even though they are of course a consequence of 
our particular way of constructing a superposition.

A visible would-be Dirac string presents a formidable obstacle to considering 
the constituents as independent objects; they remember to which caloron they 
belong. Insisting the abelian field far from the constituents to be exactly 
additive under the approximate superposition leaves us little room for 
considering other possible superpositions, apart from carefully fine-tuning 
the charged components of the configuration. Solving for the exact self-dual 
caloron solutions of higher charge we wish to show that a visible Dirac string 
is an artifact of the particular procedure to construct approximate caloron 
solutions. For the exact charge $k$ caloron solutions we expect $kn$ 
constituents and from the point of view of the parameter space these can 
be expected to be independent (as long the constituents do not get too 
close together). 

\begin{figure}[htb]
\vskip3.0cm
\includegraphics{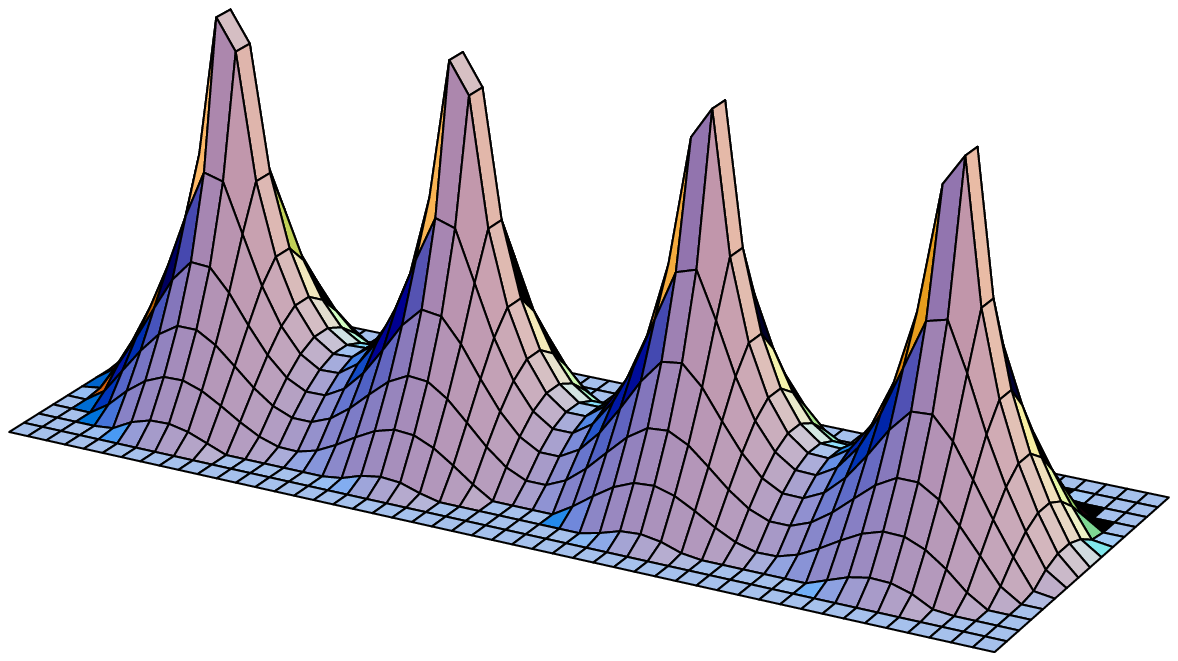}
\includegraphics{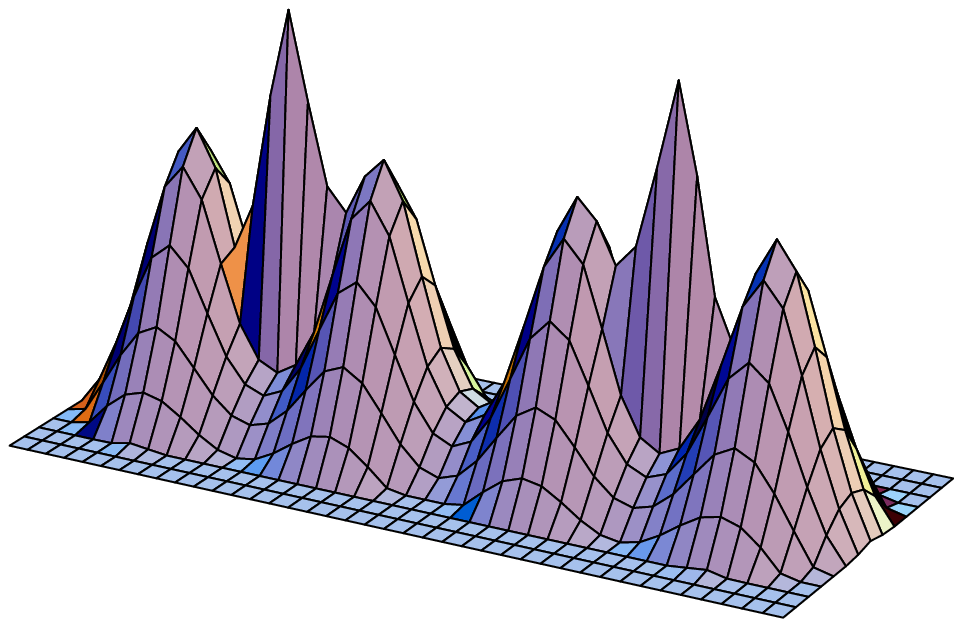}
\includegraphics{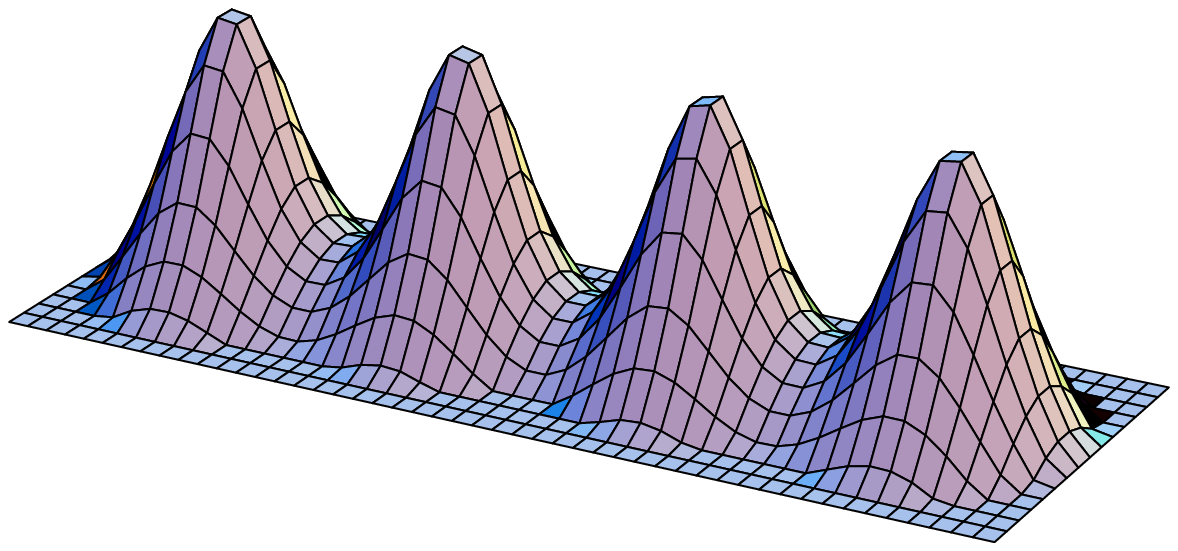}
\caption{Comparing the logarithm of the action density (cutoff for $\log(S)$ 
below -3 and above 7) as a function of $x$ and $z$ for the exact $SU(2)$ 
solution ($\mu_2=\quarter$) with charge 2 (left) with the approximate 
superposition of two charge one calorons (middle) and the abelian solution 
based on Dirac monopoles (right), all on the same scale. The two pairs of 
constituent monopoles are located at $\vec x=(0,0,6.031),~(0,0,2.031)$ and 
$\vec x=(0,0,-2.031),~(0,0,-6.031)$.}\label{fig:Wellsep}
\end{figure}
In this paper we develop the formalism and give exact solutions for a 
class of axially symmetric solutions. For $SU(2)$ an example of such an exact 
solution with charge 2 is presented in Fig.~\ref{fig:Wellsep}. We compare 
the exact solution (left), with the one obtained by adding two charge 1 
calorons, showing the would-be Dirac strings (middle) and with the exact 
abelian solution determined from (self-dual) Dirac monopoles placed at the 
location of the constituents (right). Indeed, the would-be Dirac strings 
are no longer visible for the exact non-abelian solution. This gives us 
good reasons to expect that moving away from the requirement of axial 
symmetry, the exact solutions will exhibit the constituent monopoles as 
independent objects. This is an important prerequisite for attempting to 
formulate the long distance features of QCD in terms of these monopoles. 
Non-linearities will remain important when constituents overlap, like for 
instantons at zero-temperature, as will be illustrated through suitable 
examples.

Our results can also be used to extract information on multi-monopole
solutions. In the charge 1 case it has been shown~\cite{KvB2} that sending 
one of the constituents to infinity one is left with exact monopole solutions.
Perhaps somewhat surprisingly, it has been notoriously difficult to construct 
approximate superpositions of magnetic monopoles in such a way the interaction 
energy decreases with their separation. Within our approach this is due to 
the difficulty (particularly when sending some constituents to infinity) in 
keeping Dirac strings hidden. It would therefore still be important to find 
approximate superpositions that achieve this. Nevertheless, it shows how 
subtle it is to consider abelian fields as embedded in non-abelian ones. 
There is much more to 't Hooft's abelian projection~\cite{AbPr} than meets 
the eye at first glance.

The rest of the paper will be organized as follows. In Sect.~2 we
formulate the Fourier analysis of the ADHM data for calorons of higher
charge. This allows us in Sect.~2.1 to relate to the Nahm equations,
providing the connection between the ADHM and Nahm data. The Nahm equation
for charge $k$ calorons~\cite{NCal} expresses self-duality of dual $SU(k)$
gauge fields on a circle, but with singularities. In Sect.~2.2 we introduce
the ``master'' Green's function in terms of which the gauge field (Sect.~2.3)
and the action density (Sect.~2.4) can be explicitly computed, relying 
heavily on some beautiful results~\cite{Temp,Osb} derived in the context 
of the ADHM construction.

This can all be achieved without explicitly solving the Nahm equation.
In Sect.~3 we will, however, solve it for some special cases with axial
symmetry. There we also illustrate some interesting features that 
appear when constituents overlap, similar to what was observed at
zero temperature~\cite{GaKo}. 

Sect.~4 is devoted to the far-field limit (related to the high-temperature 
limit), where up to exponential corrections only the abelian component of the 
field survives. Again, expressions can be derived without explicitly solving 
the Nahm equation. We apply the formalism to the exact axially symmetric 
solutions in Sect.~4.3.1, where we prove that in the far-field region they 
are in a precise sense described by (self-dual) Dirac monopoles, which is 
conjectured to be the case for any exact solution. 

We end in Sect.~5 with a discussion of lattice results that have been
obtained by various groups, some open questions and suggestions for
future studies.

\section{Construction}

In constructing the higher charge caloron solutions similar steps are 
followed as for charge 1. We distinguish two constructions, $Sp(n)$ based on 
quaternions, and $SU(n)$ based on complex spinors. With $SU(2)=Sp(1)$, this 
implies that for $SU(2)$ two constructions are possible. For $k=1$ it was 
easily shown these are identical, but for $k>1$ this is more complicated due 
to the quadratic ADHM constraint~\cite{ADHM} which can not be 
solved in all generality. Since the essential part of the construction 
that employs the impurity scattering technique is identical in both 
approaches, we will give a unified presentation.

The $SU(n)$ ADHM formalism for charge $k$ instantons~\cite{ADHM} 
employs a $k$ dimensional vector $\lambda=(\lambda_1,\ldots,\lambda_k)$, where 
$\lambda_i^\dagger$ is a two-component spinor in the $\bar n$ representation 
of $SU(n)$. Alternatively, $\lambda$ can be seen as a $n\times 2k$ complex 
matrix.  In addition one has four complex hermitian $k\times k$ matrices 
$B_\mu$, combined into a $2k\times2k$ complex matrix $B=\sigma_\mu\otimes 
B_\mu$, using the unit quaternions $\sigma_\mu=(\ein_2,i\vec\tau)$ and $\bar
\sigma_\mu=(\ein_2,-i\vec\tau)$, where $\tau_i$ are the Pauli matrices. With 
some abuse of notation, we often write $B=\sigma_\mu B_\mu$. Together $\lambda$
and $B$ constitute the $(n + 2 k)\times 2k$ dimensional matrix $\Delta(x)$, 
to which is associated a complex $(n+2k)\times n$ dimensional normalized 
zero mode vector $v(x)$,
\beq
\Delta(x)=\left(\!\!\bea{c}\lambda\\B(x)\eea\!\!\right),\quad B(x)=B-x,\quad
\Delta^\dagger(x)v(x)=0,\quad v^\dagger(x)v(x)=\ein_n.
\eeq
Here the quaternion $x=x_\mu\sigma_\mu$ denotes the position (a $k\times k$ 
unit matrix is implicit) and $v(x)$ can be solved explicitly in terms of the
ADHM data by
\beq
v(x)=\left(\!\bea{c}-\ein_n\\u(x)\eea\!\right)\phi^{-\half},\quad u(x)=(
B^\dagger-x^\dagger)^{-1}\lambda^\dagger,\quad\phi(x)=\ein_n+u^\dagger(x)u(x),
\eeq
As $\phi(x)$ is an $n\times n$ positive hermitian matrix, its square root 
$\phi^{\half}(x)$ is well-defined. The gauge field is given by
\beq
A_\mu(x)=v^\dagger(x)\partial_\mu v(x)=\phi^{-\half}(x)(u^\dagger(x)
\partial_\mu u(x))\phi^{-\half}(x)+\phi^{\half}(x)\partial_\mu\phi^{-\half}(x).
\label{eq:aadhm}
\eeq
The $Sp(1)$ ADHM formalism for charge $k$ instantons~\cite{ADHM} 
employs also a $k$ dimensional vector $\lambda=(\lambda_1,\ldots,\lambda_k)$, 
where now $\lambda_i$ is a quaternion (4 real parameters). Again, $\lambda$ can
be seen as an $2\times 2k$ complex matrix (but with $4k$ real, as opposed to 
complex, parameters). Now the four $k\times k$ matrices $B_\mu$ are required to
be real and symmetric, still to be combined into a $2k\times2k$ complex matrix 
$B=\sigma_\mu B_\mu$. The $(2 + 2 k)\times 2k$ dimensional matrix $\Delta(x)$ 
is constructed as before. It is immediately obvious that now $\phi(x)$ is 
proportional to $\sigma_0$, simplifying the expression for the gauge field to
\beq
A_\mu(x)=v^\dagger(x)\partial_\mu
v(x)=(u^\dagger(x)\partial_\mu u(x))/\phi(x).
\eeq

For $A_\mu(x)$ to be a self-dual connection, $\Delta(x)$ has to satisfy the 
quadratic ADHM constraint, which states that $\Delta^\dagger(x)\Delta(x)=
B^\dagger(x)B(x)+\lambda^\dagger\lambda$ (considered as a $k\times k$ complex 
quaternionic matrix) has to commute with the quaternions, or equivalently
\beq
\Delta^\dagger(x)\Delta(x)=\sigma_0\otimes f^{-1}_x,\label{eq:adhmconstr}
\eeq
defining $f_x$ as a hermitian (resp. symmetric) $k\times k$ Green's function. 
The \SD ity follows by computing the curvature
\beq
F_{\mu\nu}=2\phi^{-\half}(x)u^\dagger(x)\eta_{\mu\nu}f_x u(x)\phi^{-\half}(x),
\eeq
making essential use of the fact that $f_x$ commutes with the quaternions, and
$\eta_{\mu\nu}\equiv\sigma_{[\mu}\bar\sigma_{\nu]}$ being \SD\ ($\bar\eta_{\mu
\nu}\equiv\bar\sigma_{[\mu}\sigma_{\nu]}$ is \ASD). The quadratic constraint
can be formulated as $\myIm(\Delta^\dagger(x)\Delta(x))=0$, where
$\myIm W\equiv W-\half\sigma_0\tr_2 W$, and one obtains
\beq
\bar\eta_{\mu\nu}\otimes B_\mu B_\nu+\half\tau_a
\otimes\tr_2(\tau_a\lambda^\dagger\lambda)= 0,\label{eq:radhmcon}
\eeq
where $\tr_2$ is the spinorial trace. Note that this implies that 
$\tr_2(\tau_a\lambda^\dagger\lambda)$ is traceless for $a=1,2,3$. 

To count the number of instanton parameters we observe that the transformation 
$\lambda\rightarrow\lambda T^\dagger$, $B_\mu\rightarrow T B_\mu T^\dagger$,
with $T \in U(k)$ (resp. $T\in O(k)$) leaves the gauge field and the ADHM 
constraint untouched. Taking this symmetry into account, one checks the
dimension of the instanton moduli space to be $4kn$ dimensional.  We have $4kn$
and $4k^2$ real parameters from $\lambda$ and $B_\mu$. The $U(k)$ symmetry 
removes $k^2$ real parameters and finally the quadratic ADHM constraint 
gives $3k^2$ real equations. On the other hand, for the quaternionic
construction there are $4k$ and $4\cdot\half k(k+1)$ parameters from $\lambda$ 
and $B_\mu$, of which $\half k(k-1)$ are removed by the $O(k)$ symmetry, and 
the quadratic ADHM constraint gives $3\cdot\half k(k-1)$ equations. Global 
gauge transformations are realized by $\lambda \rightarrow g \lambda$, with 
$g\in SU(n)$ and are here included in the parameter count. For calorons with 
non-trivial holonomy the dimension of the gauge invariant parameter space is 
minimally reduced by $n-1$ (maximal symmetry breaking) and maximally by 
$n^2-1$ (trivial holonomy). 

We note that for $SU(2)$ the quadratic ADHM constraint and symmetry of the ADHM
data for high charge differ considerably. For the caloron we have to deal in a 
sense with infinite topological charge, but finite within each (imaginary) 
time interval of length $\beta$. This infinity is resolved by Fourier 
transformation, relating it to the Nahm formalism~\cite{NCal}, but the 
difference between the $U(k)$ and $O(k)$ symmetries (the infinity is moved 
to making these gauge symmetries local) remains, as well as of course the 
nature of the Nahm data (the Fourier transformation of the ADHM data). 
Henceforth we put $\beta=1$, which can always be achieved by a rescaling.

Like for charge one~\cite{KvB2,KvBn} the caloron with Polyakov loop $\pl$ at 
infinity is built out of a periodic array of instantons, twisted by $\pl$. 
This is implemented in the ADHM formalism by requiring (suppressing color 
and spinor indices, respectively quaternion indices)
\beq
u_{pk+k+a}(x+1)=u_{pk+a}(x)\cP_\infty^{-1} \label{eq:ucocyc}
\eeq
with $p\in\zahlen$ (the Fourier index), and $a=1,\ldots,k$ (associated to
the non-Abelian nature of the Nahm data). Using that $\phi^{\pm\half}(x+1)=
\pl\phi^{\pm\half}(x)\cP_\infty^{-1}$, \refeq{aadhm} leads to the required 
periodicity. Demanding
\beq
\lambda_{pk+k+a}=\pl\lambda_{pk+a},\quad B_{pk+a,qk+b}=B_{pk-k+a,qk-k+b}+
\sigma_0\delta_{pq}\delta_{ab},
\eeq
suitably implements \refeq{ucocyc} and is partially solved by imposing 
\beq
\lambda_{pk+a}=\cP^p_\infty\zeta_a, \quad B_{pk+a,qk+b}=p\sigma_0\delta_{pq}
\delta_{ab}+\hat A^{ab}_{p-q},
\eeq
with $\hat A$ still to be determined to account for \refeq{radhmcon}. It is 
useful to introduce the $n$ projectors $P_m$ on the $m$th eigenvalue of 
$\pl$, such that $\pl=\sum_m e^{2\pi i\mu_m}P_m$ and $\lambda_{pk+a}=
\sum_m e^{2\pi ip\mu_m}P_m\zeta_a$.

\subsection{Nahm setting}

We now perform the Fourier transformation to the Nahm setting~\cite{NCal},
which casts $B$ into a Weyl operator and $\lambda^\dagger\lambda$ into a 
singularity structure on $S^1$,
\beqa
&&\sum_{p,q} B_{pk+a,qk+b}(x) e^{2\pi i(pz-qz')}=\frac{\delta(z-z')}{2\pi i}
\hat D^{ab}(z'),\quad2\pi i\sum_p e^{2\pi ipz}\hat A^{ab}_{p}=\hat A^{ab}(z),
\nonumber\\ &&\sum_{p,q}\lambda^\dagger_{pk+a} e^{2\pi i(pz-qz')}\lambda^{
\phantom{\dagger}}_{qk+b}=\delta(z-z')\hat\Lambda_{ab}(z),\quad\sum_p e^{-2\pi 
ipz}\lambda_{pk+a}=\hat\lambda_a(z),
\eeqa
With $B=\sigma_\mu B^\mu$, we can write $\hat D(z)=\sigma_\mu\hat 
D^\mu(z)$ and $\hat A(z)=\sigma^\mu\hat A_\mu(z)$, where
\beqa
&&\hat D^{ab}_x(z)\equiv\hat D^{ab}(z)-2\pi ix\delta^{ab}=
\delta^{ab}(\sigma_0\ddz-2\pi ix)+\hat A^{ab}(z),\\
&&\hat\Lambda_{ab}(z)=\sum_m\delta(z-\mu_m)\zeta_a^\dagger P_m\zeta_b,
\quad\hat\lambda_a(z)=\sum_m\delta(z-\mu_m)P_m\zeta_a.\nonumber
\eeqa
It should be noted that $B_\mu$ hermitian, implies that $\hat A_\mu(z)$
is hermitian (as a $k\times k$ matrix), whereas a real symmetric $B_\mu$
in addition implies $\hat A_\mu^t(z)=\hat A_\mu(-z)$.

The $2\times 2$ matrix $\hat\Lambda^{ab}$ can always be decomposed as
\beq
\zeta_a^\dagger P_m\zeta^{\hphantom{\dagger}}_b=\frac{1}{2\pi}(\sigma_0
\hat S^{ab}_m-\vec\tau\cdot\vec\rho^{\,ab}_m).\label{eq:Lam}
\eeq
On the diagonal one can show, as for $k=1$, that $\hat S_m^{aa}=|\vec
\rho^{\,aa}_m|$, but in general the relation between $\vec \rho_m$ and 
$\hat S_m$ is more complicated,
\beq
\hat S_m^{ab}\hat S_m^{cd}+\hat S_m^{ad}\hat S_m^{cb}=\vec\rho_m^{\,ab}
\cdot\vec\rho_m^{\,cd}+\vec\rho_m^{\,ad}\cdot\vec\rho_m^{\,cb}.
\eeq
Furthermore, $\tr_2(\vec\tau\lambda^\dagger\lambda)$ is traceless implies 
that $\sum_{a=1}^k\sum_{m=1}^n\vec\rho_m^{\,aa}=\vec 0$. Both these conditions 
are equally valid for the $Sp(1)$ construction. But in the latter case, 
since through the Nahm equation $\vec\rho_m$ determines the discontinuities
in $\hat A(z)$, compatibility with $\hat A_\mu^{\,t}(z)=\hat A_\mu(-z)$ 
requires $\vec\rho_1+\vec\rho_2^{\ t}=\vec 0$, which will be verified below 
\refeq{fsymm}. The Nahm equation is obtained by Fourier transforming the 
quadratic ADHM constraint 
\beq
\half[\hat D_\mu(z),\hat D_\nu(z)]\,\bar\eta_{\mu\nu}=4\pi^2\myIm
\hat\Lambda(z),
\eeq
or in more familiar form (as a $k\times k$ matrix equation)
\beq
\ddz\hat A_j(z)+[\hat A_0(z),\hat A_j(z)]+\half\veps_{jk\ell}[\hat A_k(z),\hat 
A_\ell(z)]=2\pi i\sum_m\delta(z-\mu_m)\rho_m^{\,j},\label{eq:nahm}
\eeq
\setlength{\unitlength}{0.91cm}
\thinlines
\begin{picture}(10,1.5)(-1.6,-1.0)
\put (0.,0){$\hat A_j:$}
\put (1.2,0){\line(1,0){0.2}}
\put (1.6,0){\line(1,0){0.2}}
\put (1.25,-0.5){$z=$}
\put (2,0){\line(1,0){3}}
\put (2,-0.25){\line(0,1){0.5}}
\put (2,0){\circle*{0.1}}
\put (2,-0.5){$\mu_1$}
\put (1.7,0.5){$\rho^j_1$}
\put (3.75,-0.25){\line(0,1){0.5}}
\put (3.75,0){\circle*{0.1}}
\put (3.45,-0.5){$\mu_2$}
\put (3.45,0.5){$\rho^j_2$}
\put (5,-0.25){\line(0,1){0.5}}
\put (5,0){\circle*{0.1}}
\put (4.7,-0.5){$\mu_3$}
\put (4.7,0.5){$\rho^j_3$}
\put (5.2,0){\line(1,0){0.2}}
\put (5.6,0){\line(1,0){0.2}}
\put (6.0,0){\line(1,0){0.2}}
\put (6.4,0){\line(1,0){0.2}}
\put (6.8,0){\line(1,0){0.2}}
\put (7.25,0){\line(1,0){0.2}}
\put (7.7,0){\line(1,0){0.2}}
\put (8.1,0){\line(1,0){0.2}}
\put (8.5,0){\line(1,0){2.5}}
\put (8.5,-0.25){\line(0,1){0.5}}
\put (8.5,0){\circle*{0.1}}
\put (7.8,-.5){$\mu_{n-2}$}
\put (7.8,.5){$\rho^j_{n-2}$}
\put (9.5,-0.25){\line(0,1){0.5}}
\put (9.5,0){\circle*{0.1}}
\put (9.2,-.5){$\mu_{n-1}$}
\put (9.2,.5){$\rho^j_{n-1}$}
\put (11.0,0){\circle*{0.1}}
\put (11.,-0.25){\line(0,1){0.5}}
\put (10.7,-.5){$\mu_n$}
\put (10.7,.5){$\rho^j_n$}
\put (11.0, 0){\line(1,0){1.0}}
\put (12.0,-0.25){\line(0,1){0.5}}
\put (12.0,0){\circle*{0.1}}
\put (11.6,-.5){$\mu_{n+1}\!\equiv\! \mu_1+1$}
\put (11.7,.5){$\rho^j_{1}$}
\put (12.2, 0.){\line(1,0){0.2}}
\put (12.6, 0.){\line(1,0){0.2}}
\put (14.3,0.){.}
\end{picture}
\setlength{\unitlength}{1.0pt}\\
where the figure illustrates the jumps of $\hat A_j$ at each of the 
singularities (an overall factor of $2\pi i$ is not shown). The $T$ 
symmetry in the ADHM construction translates into a $U(k)$ gauge symmetry 
on $S^1$, which allows one to set $\hat A_0=2\pi i\xi_0$ to be constant, 
where $\xi_0$ is a hermitian matrix which can be made diagonal by a 
constant gauge transformation. Its trace part can be absorbed in $x_0$. 

\subsection{Green's function}

Central to the ADHM construction is the Green's function $f_x$, which when 
Fourier transformed to $\hat f^{ab}_x(z,z')\equiv\sum_{p,q}f_x^{pk+a,qk+b}
e^{2\pi i(pz-qz')}$ is a solution to the differential equation 
\beq
\left\{\left(\frac{\hat D_x^\mu(z)}{2\pi i}\right)^{\!\!2}\!+
\frac{1}{2\pi}\sum_m\delta(z\!-\!\mu_m)\hat S_m\right\}\!\hat f_x(z,z')=
\ein_k\delta(z\!-\!z'),\label{eq:df}
\eeq
\setlength{\unitlength}{0.91cm}
\thinlines
\begin{picture}(10,2.0)(-1.5,-1.1)
\put (0.,0.){$\frac{d\hat f_x}{dz}:$}
\put (1.2,0){\line(1,0){0.2}}
\put (1.6,0){\line(1,0){0.2}}
\put (1.25,-0.5){$z=$}
\put (2,0){\line(1,0){3}}
\put (2,-0.25){\line(0,1){0.5}}
\put (2,0){\circle*{0.1}}
\put (2,-0.5){$\mu_1$}
\put (1.7,0.5){$\hat S_1$}
\put (3.75,-0.25){\line(0,1){0.5}}
\put (3.75,0){\circle*{0.1}}
\put (3.45,-0.5){$\mu_2$}
\put (3.45,0.5){$\hat S_2$}
\put (5,-0.25){\line(0,1){0.5}}
\put (5,0){\circle*{0.1}}
\put (4.7,-0.5){$\mu_3$}
\put (4.7,0.5){$\hat S_3$}
\put (5.2,0){\line(1,0){0.2}}
\put (5.6,0){\line(1,0){0.2}}
\put (6.0,0){\line(1,0){0.2}}
\put (6.4,0){\line(1,0){0.2}}
\put (6.5,-0.25){\line(0,1){0.5}}
\put (6.5,0){\circle*{0.1}}
\put (6.1,.5){$2\pi\ein_k$}
\put (6.5,-0.5){$z'$}
\put (6.8,0){\line(1,0){0.2}}
\put (7.25,0){\line(1,0){0.2}}
\put (7.7,0){\line(1,0){0.2}}
\put (8.1,0){\line(1,0){0.2}}
\put (8.5,0){\line(1,0){2.5}}
\put (8.5,-0.25){\line(0,1){0.5}}
\put (8.5,0){\circle*{0.1}}
\put (7.8,-.5){$\mu_{n-2}$}
\put (7.8,.5){$\hat S_{n-2}$}
\put (9.5,-0.25){\line(0,1){0.5}}
\put (9.5,0){\circle*{0.1}}
\put (9.2,-.5){$\mu_{n-1}$}
\put (9.2,.5){$\hat S_{n-1}$}
\put (11.0,0){\circle*{0.1}}
\put (11.,-0.25){\line(0,1){0.5}}
\put (10.7,-.5){$\mu_n$}
\put (10.7,.5){$\hat S_n$}
\put (11.0, 0){\line(1,0){1.0}}
\put (12.0,-0.25){\line(0,1){0.5}}
\put (12.0,0){\circle*{0.1}}
\put (11.6,-.5){$\mu_{n+1}\!\equiv\! \mu_1+1$}
\put (11.7,.5){$\hat S_{1}$}
\put (12.2, 0.){\line(1,0){0.2}}
\put (12.6, 0.){\line(1,0){0.2}}
\put (14.3,0.){.}
\end{picture}
\setlength{\unitlength}{1.0pt}\\
where the figure illustrates the jumps for the {\em derivative} of $\hat f_x$ 
(itself being continuous) at each of the singularities, which now includes 
$z=z'$ (an overall factor of $2\pi$ is omitted). With help of the impurity 
scattering formalism we will be able to express the solution in a simple form, 
without assuming explicit knowledge of $\hat A(z)$. As a bonus this also 
gives a more transparent derivation for $k=1$. The equation for the Green's 
function is valid for the $SU(n)$ and $Sp(1)$ formulations alike. In general 
the matrix $f_x$ is complex hermitian, but for $Sp(1)$ it is real symmetric, 
which implies that
\beq
\hat f_x^{ab}(z,z')=\hat f_x^{ba}(z',z)^*,\quad\mbox{and for 
$Sp(1)$ only}\quad\hat f_x^{ab}(z,z')=\hat f_x^{ba}(-z',-z).
\label{eq:fsymm}
\eeq
For $Sp(1)$ consistency requires, $\hat S_1-\hat S_2^{\,t}=0$, which is to be 
compared with the condition we found in the previous section, $\vec\rho_1+\vec 
\rho_2^{\,t}=0$. To demonstrate the validity of these relations we make use of 
the fact that $\zeta_a=\zeta_a^\mu\sigma_\mu$, with $\zeta_a^\mu$ real (such 
that $\zeta_a^\dagger=\bar\zeta_a$). Furthermore, we note that with $\pl\equiv
\exp(2\pi i\vec\omega\cdot\vec\tau)$, $\mu_2=-\mu_1=\omega\equiv|\vec\omega|$, 
whereas $P_1=\half(\ein_2-\hat\omega\cdot\vec\tau)$ and $P_2=\half(\ein_2+
\hat\omega\cdot\vec\tau)$. Introducing $\Omega\equiv\hat\omega\cdot
\vec\sigma=i\hat\omega\cdot\vec\tau$, which is an imaginary unit quaternion 
(i.e. $\bar\Omega=-\Omega$, $\bar\Omega\Omega=\sigma_0$), we find with the 
help of \refeq{Lam} 
\beqa
\sigma_0\hat S^{ab}_m-\vec\tau\cdot\vec\rho^{\,ab}_m&=&\pi\zeta_a^\mu\bar
\sigma_\mu(\sigma_0-i(-1)^m\Omega)\zeta^{\nu}_b\sigma_\nu\label{eq:rhoS}\\
&=&\pi\zeta_a^\mu\zeta_b^\nu\left(\left[\bar\sigma_{\{\mu}\sigma_{\nu\}}-
i(-1)^m\bar \sigma_{[\mu}\Omega\sigma_{\nu]}\right]+\left[\bar\sigma_{[\mu}
\sigma_{\nu]}-i(-1)^m\bar\sigma_{\{\mu}\Omega\sigma_{\nu\}}\right]\right)
\nonumber\\&=&\sigma_0\pi\left[\zeta_a^\mu\zeta_b^\mu-i(-1)^m\hat\omega\cdot
\vec\eta_{\mu\nu}\zeta_a^\mu\zeta_b^\nu\right]+\pi\left[\bar\eta_{\mu\nu}
\zeta_a^\mu\zeta_b^\nu+(-1)^m\bar\zeta_{\{a}\hat\omega\cdot\vec\tau
\zeta_{b\}}\right],\nonumber
\eeqa
for which we used that $\bar\sigma_{[\mu}\Omega\sigma_{\nu]}$ is a real
quaternion, therefore given by $\half\sigma_0\tr_2(\bar\sigma_{[\mu}\Omega
\sigma_{\nu]})=\half\sigma_0\tr_2(\Omega\eta_{\nu\mu})=\sigma_0\hat\omega
\cdot\vec\eta_{\mu\nu}$. We directly read off $\hat S_m$ and $\vec\rho_m$ 
and verify that indeed $\hat S_1=\hat S_2^{\,t}$ and $\vec\rho_1=-\vec 
\rho_2^{\,t}$. We wrote the result so as to easily make contact with the 
$k=1$ results~\cite{KvB2} (for which the 2nd term in $\hat S_m$ and the 
1st term in $\vec\rho_m$ do not contribute).

In formulating the impurity scattering we note that the $U(k)$ gauge 
transformation 
\beq
\hat g(z)=\exp\left(2\pi i(\xi_0-x_0\ein_k)z\right)
\eeq
turns $\hat D_x^0(z)$ into the ordinary derivative and conjugates all other 
objects in \refeq{df}, such that
\beq
\left\{-\frac{d^2}{dz^2}+V(z;\vec x)\right\}
\!f_x(z,z')=4\pi^2 \ein_k\delta(z\!-\!z'),\label{eq:dftrans}
\eeq
with $f_x(z,z')$ and $V(z;\vec x)$ given by
\beq
f_x(z,z')\equiv\hat g(z)\hat f_x(z,z')\hat g^\dagger(z'),\quad V(z;\vec x)
\equiv4\pi^2\vec R^2(z;\vec x)+2\pi\sum_m\delta(z-\mu_m)S_m,\label{eq:ftofh}
\eeq
where $\vec R(z;\vec x)$ and $S_m$ are defined by
\beq
R_j(z;\vec x)\equiv x_j\ein_k-\frac{1}{2\pi i}\hat g(z)\hat A_j(z)\hat g^\dagger
(z),\quad S_m\equiv\hat g(\mu_m)\hat S_m\hat g^\dagger(\mu_m).\label{eq:Rdef}
\eeq 
Periodicity is now only up to a gauge transformation. Note that $R_j(z;\vec x)$
is a hermitian $k\times k$ matrix, and that $V(z;\vec x)$ does {\em not} depend
on $x_0$. By combining $f_x$ and its derivative into a vector,
\beq
\vec f_x(z,z')\equiv\pmatrix{\quad f_x(z,z')\cr\ddz f_x(z,z')\cr},
\eeq
we can turn the second order equation in to a first order equation
\beq
\left\{\frac{d}{dz}-\pmatrix{0&\ein_k\cr V(z;\vec x)&0\cr}\right\}\vec f_x
(z,z')=\vec C\delta(z\!-\!z'),\quad\vec C\equiv\:-4\pi^2\pmatrix{0\cr\ein_k\cr}
\eeq
Its solution is
\beq
\vec f_x(z,z')=W(z)\left[\vec c(z')-\theta(z'-z)W^{-1}(z')\vec C\right],
\quad\ddz W(z)=\pmatrix{0&\ein_k\cr V(z;\vec x)&0\cr}W(z)
\eeq
where $W$ is a $2\times 2$ matrix, whereas $\vec c$ is a two-component vector
(like $\vec C$), all components being hermitian $k\times k$ matrices. The 
theta function takes care of the inhomogeneous part of the equation. However, 
we have to restrict to $z'\in[z-1,z+1]$ since the delta function is periodic. 
The solution can be extended beyond this range using the periodicity, which 
when imposed, as we will see, also determines $\vec c(z')$.

The solution for $W(z)$ can be (formally) written as a path ordered exponential
integral in the usual way. To show that this indeed makes sense, we should 
specify how to deal with the delta functions in $V$. From the definition of 
the path ordered exponential integral we find $W(\mu_m+0)=T_m W(\mu_m-0)$ with
\beq
T_m\equiv\Pexp\left[\int_{\mu_m-0}^{\mu_m+0}\pmatrix{0&\ein_k\cr 
V(z;\vec x)&0\cr}dz\right]=\exp\pmatrix{0&0\cr 2\pi S_m&0\cr}= 
\pmatrix{\ein_k&0\cr 2\pi S_m&\ein_k\cr},
\eeq
which correctly reflects the matching conditions due to the ``impurity" at 
$z=\mu_m$, since $f_x(z,z')$ is continuous across $z=\mu_m$, whereas the 
derivative jumps with $2\pi S_m f_x(\mu_m,z')$ and both conditions can
be summarized by $\vec f_x(\mu_m+0,z')=T_m\vec f_x(\mu_m-0,z')$. Note that 
$\vec f_x$ evolves with $z$ as $\vec f_x(z_2,z')=W(z_2)W^{-1}(z_1)
\vec f_x(z_1,z')$, with 
\beq
W(z_2)W^{-1}(z_1)=\Pexp\left[\int_{z_1}^{z_2}\pmatrix{0&\ein_k\cr
V(z;\vec x)&0\cr}dz\right]\equiv W(z_2,z_1).\label{eq:Wdef}
\eeq
In particular, when $z_1=\mu_m$ and $z_2=\mu_{m+1}$ this gives the 
``propagation" between two neighboring impurities and we can write
\beq
H_m\equiv W(\mu_{m+1}-0,\mu_m+0)=\Pexp\left[\int_{\mu_m}^{\mu_{m+1}}
\pmatrix{0&\ein_k\cr 4\pi^2\vec R^2(z;\vec x)&0\cr}dz\right].\label{eq:Hm}
\eeq

Neither $T_m$ nor $H_m$ require us to specify a boundary condition for $W(z)$. 
A change in boundary condition, however, affects $\vec c(z')$. To avoid such 
ambiguities, we define $\vec c_{z_0}(z')\equiv W(z_0)\vec c(z')$ such that
\beq
\vec f_x(z,z')=W(z,z_0)\left[\vec c_{z_0}(z')-\theta(z'-z)W^{-1}(z',z_0)\vec C
\right].\label{eq:fxz0}
\eeq
We determine $\vec c_{z_0}(z')$ by scattering ``around" the circle determined 
from the boundary conditions of $\vec f_x$. Using \refeq{ftofh}, the fact 
that $\hat f_x(z,z')$ is strictly periodic and that $\hat g(z+1)=\hat g(1)\hat 
g(z)$, one finds 
$\vec f_x(z,z'+1)=\vec f_x(z,z')\hat g^\dagger(1)$ and $\vec f_x(z+1,z')=
\hat g (1)\vec f_x(z,z')$ (where the order of the {\em matrix} multiplication,
defined componentwise, is important). The latter condition can be used to fix 
$\vec c_{z_0}(z')$ over the range of one period, $z'\in[z,1+z]$ (the use of 
$\hat f_x(z,z')$ requires $z'\in[z-1,z+1]$, that of $\hat f_x(z+1,z')$ further 
restricts the range for $z'$ to $z'>z$). With the help of the periodicity 
properties of $W$ it gives $\vec c_{z_0}(z')=\left(\ein_{2k}-\hat g^\dagger(1)
W(z_0+1,z_0)\right)^{-1}W^{-1}(z',z_0)\vec C$. We thus find for {\em arbitrary}
$z_0$
\beq
\vec f_x(z,z')=W(z,z_0)\left\{(\ein_{2k}-\cF_{z_0})^{-1}-\theta(z'-z)\ein_{2k}
\right\}W^{-1}(z',z_0)\vec C,\label{eq:vecf}
\eeq
where we introduced the ``holonomy" $\cF_{z_0}$
\beq
\cF_{z_0}\equiv\hat g^\dagger(1)W(z_0+1,z_0)=g^\dagger(1)\Pexp
\left[\int_{z_0}^{1+z_0}\pmatrix{0&\ein_k\cr V(z;\vec x)&0\cr}dz\right].
\eeq
The equation for $\vec f_x$ is valid for $z'\in[z,z+1]$, but can be extended 
with the appropriate periodicity specified above. We note that $\hat 
g^\dagger(1)$ plays the role of a cocycle, in the gauge where $\hat A_0(z)
-2\pi ix_0\ein_k$ is transformed to 0, with $\cF_{z_0}$ the full circle 
``scattering" matrix. The $z_0$-independence of $\vec f_x$ follows from
\beqa
\cF_{z_0}&=&g^\dagger(1)W(z_0+1,z_0^\prime+1)W(z^\prime_0+1,z_0)=
W(z_0,z^\prime_0)g^\dagger(1)W(z^\prime_0+1,z_0)\nonumber\\&=&
W(z_0,z^\prime_0)\cF_{z_0^\prime}W^{-1}(z_0,z^\prime_0).\label{eq:Frel}
\eeqa
Putting things together we therefore find
\beqa
\hat f_x(z,z')&=&\hat g^\dagger(z)\left(\ein_k,0\right)\cdot\vec f_x(z,z')
\hat g(z'),\nonumber\\ \hat D_x^0(z)\hat f_x(z,z')&=&\hat g^\dagger(z)
\left(0,\ein_k\right)\cdot\vec f_x(z,z')\hat g(z'),\label{eq:finf}
\eeqa
satisfying all required conditions as can be checked explicitly.

Interestingly, \refeq{Frel} implies $\cF_{z_0+1}=\hat g(1)\cF_{z_0}\hat
g^\dagger(1)$ as should of course be the case. In the light of this we
also note that (choosing $z_0=\mu_m+0$)
\beqa
\cF_{\mu_m}&=&\hat g^\dagger(1)T_{m+n}H_{m+n-1}T_{m+n-1}H_{m+n-2}\cdots 
 T_{m+1}H_m\nonumber\\&=&T_mH_{m-1}\cdots T_2H_1T_1\hat g^\dagger(1)H_n
T_nH_{n-1}\cdots T_{m+1}H_m,\label{eq:THs}
\eeqa
\setlength{\unitlength}{0.91cm}
\thinlines
\begin{picture}(10,2.0)(-2.8,-1.1)
\put (-0.2,0.){$\cF_{\mu_m}:$}
\put (0.8,-0.5){$z=1\!+\!\mu_m$}
\put (2,0){\line(1,0){2}}
\put (2,-0.25){\line(0,1){0.5}}
\put (2,0){\circle*{0.1}}
\put (1.7,0.5){$T_m$}
\put (2.5,0.2){$H_{m-1}$}
\put (4,0){\circle*{0.1}}
\put (4,-0.25){\line(0,1){0.5}}
\put (3.4,-0.5){$1\!+\!\mu_{m-1}$}
\put (3.7,0.5){$T_{m-1}$}
\put (4.0,0){\line(1,0){0.2}}
\put (4.4,0){\line(1,0){0.2}}
\put (4.8,0){\line(1,0){0.2}}
\put (5.2,0){\line(1,0){0.2}}
\put (5.6,0){\line(1,0){0.2}}
\put (6.0,0){\line(1,0){0.2}}
\put (6.4,0){\circle*{0.1}}
\put (6.4,-0.25){\line(0,1){0.5}}
\put (5.7,.5){$T_1\hat g^\dagger(1)$}
\put (5.8,-0.5){$1\!+\!\mu_1$}
\put (6.4,0){\line(1,0){2}}
\put (7.2,0.2){$H_n$}
\put (8.4,-0.25){\line(0,1){0.5}}
\put (8.4,0){\circle*{0.1}}
\put (8.1,.5){$T_n$}
\put (8.1,-0.5){$\mu_n$}
\put (8.4,0){\line(1,0){0.2}}
\put (8.8,0){\line(1,0){0.2}}
\put (9.2,0){\line(1,0){0.2}}
\put (9.6,0){\line(1,0){0.2}}
\put (10,0){\line(1,0){2}}
\put (10,-0.25){\line(0,1){0.5}}
\put (10,0){\circle*{0.1}}
\put (9.7,-.5){$\mu_{m+1}$}
\put (9.7,.5){$T_{m+1}$}
\put (10.0,0){\circle*{0.1}}
\put (10.9,0.2){$H_m$}
\put (12.0,0){\circle*{0.1}}
\put (12.,-0.25){\line(0,1){0.5}}
\put (11.7,-.5){$\mu_m$}
\put (12.2,0.){,}
\end{picture}
\setlength{\unitlength}{1.0pt}\\
using for the second identity that $T_{m+n}=\hat g(1)T_m\hat g^\dagger(1)$ 
and $H_{m+n}=\hat g(1)H_m\hat g^\dagger(1)$. We will use these ingredients 
further on to relate to the earlier results for $k=1$, where the positioning 
of $\hat g^\dagger(1)$ is of course irrelevant.

\subsection{Gauge field}\label{sec:gaugefield}

The central role of the Green's function $\hat f_x(z,z')$ becomes clear 
when one appeals to the fact that it can be used to find the gauge field 
(working out \refeq{aadhm}), whereas its determinant gives a simple 
expression for the action density. This follows from the general ADHM 
construction~\cite{Temp,Osb}, and can be directly taken over for the 
caloron~\cite{KvB2,KvBn,Dubna}. For the gauge field one finds 
\beqa
A_\mu(x)&=&\half\phi^\hhalf(x)\lambda\bar\eta_{\mu\nu}\partial_\nu f_x
\lambda^\dagger\phi^\hhalf(x)+\half[\phi^{-\hhalf}(x),\partial_\mu\phi^\hhalf
(x)]\nonumber\\&=&\half\phi^\hhalf(x)\bar\eta_{\mu\nu}^j\partial_\nu\phi_j(x)
\phi^\hhalf(x)+\half[\phi^{-\hhalf}(x),\partial_\mu\phi^\hhalf(x)],
\label{eq:Adef}
\eeqa
where $\phi(x)$ and $\phi_j(x)$ are $n\times n$ matrices defined by
\beq
\phi(x)\equiv\left(\ein_n-\lambda f_x\lambda^\dagger\right)^{-1},
\quad\phi_j\equiv\lambda\sigma_j f_x\lambda^\dagger.
\eeq
To apply this to the caloron all we have to do is perform the Fourier 
transformation, 
\beq
\phi(x)^{-1}=\ein_n-\sum_{m,m'}P_m\zeta_a \hat f_x^{ab}(\mu_m,\mu_{m'})
\zeta^{\dagger}_bP_{m'},\quad\phi_j=\sum_{m,m'}P_m\zeta_a \sigma_j
\hat f_x^{ab}(\mu_m,\mu_{m'})\zeta^{\dagger}_bP_{m'}.\label{eq:phidef}
\eeq

For the $Sp(1)$ construction $\phi(x)$ is a real quaternion, and hence a 
multiple of $\sigma_0$, after which the gauge field simplifies to
\beq
A_\mu(x)=\half\phi(x)\bar\eta_{\mu\nu}^j\partial_\nu\phi_j(x).\label{eq:ASp1}
\eeq
To simplify $\phi(x)$ and $\phi_j(x)$ we use \refeq{fsymm}, together with 
$\mu_1=-\mu_2$, such that $\hat f_x^{ab}(\mu_1,\mu_1)=\hat f_x^{ba}(\mu_2,
\mu_2)$ and $\hat f_x^{ab}(\mu_1,\mu_2)=\hat f_x^{ba}(\mu_1,\mu_2)=\hat 
f_x^{ba}(\mu_2,\mu_1)^*$. We note that $\phi(x)$ and $\phi_j(x)$ involve the 
combinations $\zeta_a^{\hphantom{\dagger}}\sigma_\mu\zeta_b^\dagger=\zeta_a
\sigma_\mu\bar\zeta_b$, which can be split in symmetric and anti-symmetric 
combinations (cf. \refeq{rhoS}) 
\beqa
\zeta_a\sigma_0\bar\zeta_b&=&\sigma_0\zeta_a^\mu\zeta_b^\mu+\eta_{\mu\nu}
\zeta_a^\mu\zeta_b^\nu,\nonumber\\\zeta_a\sigma_j\bar\zeta_b&=&\sigma_0\bar
\eta^j_{\mu\nu}\zeta_a^\mu\zeta_b^\nu+\zeta_{\{a}\sigma_j\bar\zeta_{b\}}.
\eeqa
No contributions from $\hat f^{ab}_x(\mu_1,\mu_2)$ and $\hat f^{ab}_x(\mu_2,
\mu_1)$ can appear in $\phi(x)$ since these are symmetric in $a$ and $b$, 
selecting from $\zeta_a\sigma_0\bar\zeta_b$ the term proportional to 
$\sigma_0$, but $P_1\sigma_0P_2$ and $P_2\sigma_0P_1$ vanish. Therefore 
(cf. \refeq{rhoS})
\beqa
&\hskip-2cm\phi(x)^{-1}=\sigma_0-\hat f_x^{ab}(\mu_2,\mu_2)\sum_{m=1}^2\left(
P_m\zeta_a^\mu\zeta_b^\mu P_m+(-1)^mP_m\eta_{\mu\nu}\zeta_a^\mu\zeta_b^\nu P_m
\right)=\label{eq:phis}\\&\sigma_0\left[1-\hat f_x^{ab}(\mu_2,\mu_2)
\left(\zeta_a^\mu\zeta_b^\mu+i\hat\omega\cdot\vec\eta_{\mu\nu}\zeta_a^\mu
\zeta_b^\nu\right]\right)=\sigma_0\left[1-\pi^{-1}\Tr_k(\hat f_x(\mu_2,\mu_2)
\hat S_2)\right].\nonumber
\eeqa
Similarly we can simplify the expression for $\phi_j(x)$, which we split 
in a charged component and abelian, or neutral, component $\phi_j(x)=
\phi_j^{\rm ch}(x)+\phi_j^{\rm abel}(x)$ with
\beq
\phi_j^{\rm ch}=\hat f_x^{ab}(\mu_1,\mu_2)P_1\zeta_{\{a}\sigma_j\bar\zeta_{b\}}
P_2-\mbox{h.c.}
\eeq
and
\beqa
&\hskip-3cm\phi_j^{\rm abel}(x)=\hat f_x^{ab}(\mu_2,\mu_2)\sum_{m=1}^2
\left( P_m\zeta_{\{a}\sigma_j\bar\zeta_{b\}}P_m+(-1)^mP_m\bar\eta_{\mu\nu}^j
\zeta_a^\mu \zeta_b^\nu P_m\right)\label{eq:vphis}\\&=i\hat\omega\cdot
\vec\tau\hat f_x^{ab}(\mu_2,\mu_2)\left[\half\tr_2(\hat\omega\cdot\vec\tau
\zeta_{\{a}\tau_j\bar\zeta_{b\}})-i\bar\eta_{\mu\nu}^j\zeta_a^\mu\zeta_b^\nu
\right]=-i\hat\omega\cdot\vec\tau\pi^{-1}\Tr_k(\hat f_x(\mu_2,\mu_2)\rho_2^j),
\nonumber
\eeqa
where we used that $\half\tr_2(\hat\omega\cdot\vec\tau\zeta_{\{a}\tau_j\bar
\zeta_{b\}})=\half\tr_2(\bar\zeta_{\{b}\hat\omega\cdot\vec\tau\zeta_{a\}}
\tau_j)$ to correctly identify $\vec\rho_2^{\,ba}$, see \refeq{rhoS}. We may 
of course express $\phi(x)$ and $\phi_j^{\,\rm abel}(x)$ also in terms of the 
first impurity, $\phi^{-1}(x)=\sigma_0\left[1-\pi^{-1}\Tr_k(\hat f_x(\mu_1,
\mu_1)\hat S_1)\right]$ and $\phi_j^{\rm abel}(x)=i\hat\omega\cdot\vec\tau
\pi^{-1}\Tr_k(\hat f_x(\mu_1,\mu_1)\rho_1^j)$, as is easily verified.

Like for $k=1$ we will show further on that $\hat f_x^{ab}(\mu_1,\mu_2)$
decays exponentially, away from the cores of the constituent monopoles, where
only the abelian component survives,
\beq
A_\mu^{\rm abel}(x)=-\frac{i}{2}\hat\omega\cdot\vec\tau\left[1-\pi^{-1}\Tr_k
(\hat f_x(\mu_2,\mu_2)\hat S_2)\right]^{-1}\bar\eta^j_{\mu\nu}\partial_\nu
\left[\pi^{-1}\Tr_k(\hat f_x(\mu_2,\mu_2)\rho_2^{\,j})\right].
\eeq
Note that for $k=1$ we are able to write $A_\mu^{\rm abel}(x)=-\frac{i}{2}\hat
\omega\cdot\vec\tau e_j\bar\eta^j_{\mu\nu}\partial_\nu\log\phi(x)$, with 
$\vec e=\vec\rho_2/|\vec\rho_2|$, making use of the fact that $\hat S_m=
|\vec\rho_m|$, but that this is no longer true for higher charge, despite the 
fact that on the diagonal we still have $\hat S^{aa}_m=|\vec\rho_m^{\,aa}|$. 
This seems to allow for the dipoles of $k$ well-separated calorons to point 
in different directions. We will discuss this issue further when studying 
the limiting behavior far from the cores of the constituents, where the 
field becomes algebraic and constituent locations are readily identified. 

\subsection{Action density}

Within the ADHM formalism the action density is given by~\cite{Osb},
\beq
-\half\Tr_n F_{\mu\nu}^2(x)=-\half\partial_\mu^2\partial_\nu^2\log 
\psi(x),\label{eq:Osborn}
\eeq
where $\psi(x)$ equals $1/\det(f_x)$ after regularization to extract an
irrelevant overall and for calorons divergent constant. We will be able to 
find a simple expression for $\psi(x)$ at any $k$, generalizing the result 
for charge 1 calorons. We use that $\partial_\nu\log\det(f_x)=-\Tr(\{ 
\partial_\nu f_x^{-1}\}f_x)=\frac{1}{\pi i}\hat\Tr(\hat D_x^\nu\hat f_x)$, 
where in the last step we performed the Fourier transformation, and 
$\hat\Tr$ includes an integration with respect to $z$. The case $\nu=0$ is 
treated separately, due to the discontinuity in the $z$ derivative of 
$\hat f_x(z,z')$ at $z=z'$, which we regularize using point-splitting:
\beqa
\frac{1}{\pi i}\hat\Tr(\hat D_x^0\hat f_x)&=&\lim_{\veps\rightarrow0}\frac{
1}{2\pi i}\int_0^1\!dz~\Tr_k\left(\ddz f_x(z+\veps,z')+\ddz f_x(z-\veps,z')
\right)_{z'=z}\nonumber\\&=&4\pi i\int_0^1\!dz~\Tr\left(W^{-1}(z,z_0)
\pmatrix{0&0\cr0&\ein_k\cr}W(z,z_0)\left[(\ein_{2k}\!-\!\cF_{z_0})^{-1}
\!\!-\half\ein_{2k}\right]\right)\nonumber\\&=&2\pi i\int_0^1\!dz~\left[
\Tr\left((\ein_{2k}\!-\!\cF_{z_0})^{-1}\!\!-\half\ein_{2k}\right)-
\frac{1}{4\pi^2}\ddz\Tr_k(f_x(z,z))\right].
\eeqa
The Tr without an index or hat indicates the full trace over the $2k\times 
2k$ matrix involved. To see how the total derivative term appears (not 
contributing to the integral due to the periodicity of $\Tr_k[f_x(z,z)]$)
we use that
\beqa
\ddz\Tr_k(\hat f_x(z,z))\!=\!-4\pi^2\ddz\Tr\left(W^{-1}(z,z_0)\pmatrix{0&0\cr
\ein_k&0\cr}\!W(z,z_0)\left[(\ein_{2k}\!-\!\cF_{z_0})^{-1}\!\!-s \ein_{2k}
\right]\right)\nonumber\\=\!-4\pi^2\Tr\left(W^{-1}(z,z_0)\left[\pmatrix{0&0\cr
\ein_k&0\cr},\pmatrix{0&\ein_k\cr V(z;\vec x)&0\cr}\right]\!W(z,z_0)\left[(
\ein_{2k}\!-\!\cF_{z_0})^{-1}\!\!-s\ein_{2k}\right]\right)\nonumber\\=-4\pi^2
\Tr\left(W^{-1}(z,z_0)\pmatrix{-\ein_k&0\cr0&\ein_k\cr}W(z,z_0)\left[(\ein_{2k}
\!-\!\cF_{z_0})^{-1}\!\!-s \ein_{2k}\right]\right)\!,
\eeqa
with $s=\half$ (as for point-spitting, although one checks that the $s$ 
dependent term actually vanishes). We note that $\cF_{z_0}$ depends on 
$x_0$ only through $\hat g(1)$, and that $\partial_0\hat g(1)=2\pi i\hat 
g(1)$, such that $\partial_0\cF_{z_0}=2\pi i\cF_{z_0}$. With this we find
\beq
\partial_0\log\det(\hat f_x)=-\partial_0\log\psi,\quad\psi\equiv\det
\left(ie^{-\pi ix_0}(\ein_{2k}-\cF_{z_0})/\sqrt{2}\right),\label{eq:defpsi}
\eeq
which is {\em independent} of $z_0$. The factor $i/\sqrt{2}$ in the 
argument of the determinant was inserted just so $\psi$ agrees with 
the definition introduced earlier for $k=1$.

Next we compute $\partial_j\log\det(\hat f_x)$, 
\beqa
\frac{1}{\pi i}\int_{0}^{1}\! dz~\Tr_k(\hat D_x^j\hat f_x(z,z))=-2\int_0^1 
dz~\Tr_k\left(R_j(z;\vec x)f_x(z,z)\right)=\hskip3cm\nonumber\\8\pi^2\int_0^1
dz~\Tr\left(W^{-1}(z,z_0)\pmatrix{0&0\cr R_j(z;\vec x)&0\cr}W(z,z_0)\left[
(\ein_{2k}\!-\!\cF_{z_0})^{-1}\!\!-s\ein_{2k}\right]\right)=\hskip-4mm
\nonumber\\ 
\Tr\!\left((\ein_{2k}\!-\!\cF_{z_0})^{-1}\cF_{z_0}\int_{z_0}^{1+z_0}\!\!dz~
W^{-1}(z,z_0)\partial_j\pmatrix{0&\ein_k\cr V(z;\vec x)&0\cr}W(z,z_0)\right),
\eeqa
where again $s$ can take any value, but for convenience is best set to 1 here. 
Finally noting that $\cF_{z_0}$ only depends on $\vec x$ through $\vec R(z;
\vec x)$ and using that
\beq
W^{-1}(z,z_0)\partial_j W(z,z_0)=\int_{z_0}^{1+z_0} dz~W^{-1}(z,z_0)
\partial_j\pmatrix{0&\ein_k\cr V(z;\vec x)&0\cr}W(z,z_0),
\eeq
we verify that $\partial_\mu\log\det(\hat f_x)=-\partial_\mu\log\psi$ for all 
$\mu$ and $k$. It is amusing to note that this implies the remarkable formula
\beq
-\half\Tr_n F_{\mu\nu}^2(x)=-\half\partial_\mu^2\partial_\nu^2
\log\det\left(\ein_{2k}-\hat g^\dagger(1)\Pexp\left[\int_0^1\pmatrix{0&\ein_k
\cr V(z;\vec x)&0\cr}dz\right]\right),
\eeq
even though explicit evaluation can be quite cumbersome. Not so for some
special cases, including the single caloron $k=1$, where $\vec R(z;\vec x)$ is
{\em piecewise constant} as we will discuss next. 

\section{Special cases}

Consider $\vec R(z;\vec x)$ to be piecewise constant, for $z\in[\mu_m,
\mu_{m+1}]$ defined to be $\vec x-\vec Y_m$ (cf. \refeq{Rdef}), with 
$\vec Y_m$ constant hermitian $k\times k$ matrices related to the Nahm 
potential by $\hat A_j(z)=2\pi i\hat g^\dagger(z)Y^j_m\hat g(z)$. In 
this case we can easily deal with the path ordered exponential integrals.
To be specific, the ``propagation" from $\mu_m$ to $\mu_{m+1}$ defined 
through $H_m$ in \refeq{Hm}, is given by 
\beq 
H_m=\pmatrix{\cosh(2\pi\nu_m R_m)&(2\pi R_m)^{-1}\sinh(2\pi\nu_m R_m)\cr 
2\pi R_m\sinh(2\pi\nu_m R_m)&\cosh(2\pi\nu_m R_m)\cr},\label{eq:conHm}
\eeq
where $\nu_m=\mu_{m+1}-\mu_m$ (with $\mu_{m+n}=1+\mu_m$ such that
$\sum_m\nu_m=1$) and $R_m\equiv(\vec R_m\cdot\vec R_m)^\hhalf$ (a hermitian 
$k\times k$ matrix). Since $\cosh(y)$ and $y^{\pm 1}\sinh(y)$ are both 
quadratic in $y$, actually no square root is involved in this expression 
for $H_m$.

\subsection{The known charge 1-caloron}

For charge 1 the Nahm equation (\refeq{nahm}) has no commutator terms and 
$\vec R(z;\vec x)$ is always piecewise constant. It reduces to the $m^{\rm th}$
constituent radius for $z\in[\mu_m,\mu_{m+1}]$, $R_m=r_m\!\equiv\!|\vec r_m|$, 
whereas the prefactor at the impurity becomes $S_m=|\vec r_{m}-\vec r_{m-1}|\!
=\!|\vec\rho_{m}|$. With
\beq
\cA_m\equiv r_m^{-1}\pmatrix{r_m&|\vec\rho_{m+1}|\cr 0&r_{m+1}\cr}
\pmatrix{\cosh(2\pi\nu_mr_m)&\sinh(2\pi\nu_mr_m)\cr
\sinh(2\pi\nu_mr_m)&\cosh(2\pi\nu_mr_m)\cr},
\eeq
a link to the earlier charge 1 results~\cite{KvBn} is established by noting 
that
\beq
\cA_m=\pmatrix{0&1\cr 2\pi r_{m+1}&0\cr}T_{m+1}H_m
\pmatrix{0&1\cr 2\pi r_m&0\cr}^{-1}.
\eeq
With the placing of $\hat g^\dagger(1)$ irrelevant, and the possibility of 
absorbing $\xi_0$ in $x_0$, we therefore find 
\beq
\cF_{\mu_m}^{k=1}=\pmatrix{0&1\cr 2\pi r_m&0\cr}e^{2\pi ix_0}\cA_{m-1}\cA_{m-2}
\cdots\cA_1\cA_n\cdots\cA_{m+1}\cA_m\pmatrix{0&1\cr 2\pi r_m&0\cr}^{-1},
\eeq
cf. \refeq{THs}. In particular this shows that $\psi=-\half e^{-2\pi ix_0}
\det(\ein_2-\cF_{ \mu_m})$ agrees with the result found earlier, $\psi=\half
\tr(\cA_n\cA_{n-1}\cdots\cA_1)-\cos(2\pi x_0)$. The Green's functions can be 
shown to agree as well, using $\left(\ein_2-\cF_{\mu_m}\right)^{-1}=\left(
\ein_2-\bar\sigma_2\cF^{\,t}_{\mu_m}\sigma_2\right)/\det(\ein_2-\cF_{\mu_m})$.

\subsection{Exact axially symmetric solution}\label{sec:axial}

Arranging $\vec R(z;\vec x)$ to be piecewise constant when $k>1$ requires 
one to fulfill some constraints. To solve the Nahm equation, \refeq{nahm}, in 
terms of the $Y^j_m=\frac{1}{2\pi i}\hat g(z)\hat A_j(z)\hat g^\dagger(z)$,
the commutator term should vanish. One way to achieve this, is by choosing  
$\vec Y_m=Y_m\vec e$. The Nahm equation relates the discontinuities of 
$\hat A_j(z)$ to $\vec\rho_m$,
\beq
\hat g^\dagger(\mu_m)(Y_m-Y_{m-1})\hat g(\mu_m)\vec e=\vec\rho_m,
\label{eq:jumps}
\eeq
which imposes constraints on $\zeta_a$, see \refeq{Lam}. To seek a solution 
we choose all $\zeta^a$ to be parallel in group space, $\zeta^a=\rho_a\zeta$ 
($\rho_a$ a positive real number). This reduces the problem to $k=1$, since 
$\zeta_a^\dagger P_m\zeta_b=\rho_a\rho_b\zeta^\dagger P_m\zeta$ is 
proportional to $\zeta^\dagger P_m\zeta$. For $SU(2)$ this already solves 
the constraint, since for $k=1$ one has $\vec\rho_1=-\vec\rho_2$. For $n>2$
it has been shown~\cite{KvBn} that $\vec \rho_m$ can take any value, provided 
$\sum_{m=1}^n\vec\rho_m=\vec 0$, in particular we may choose all $\vec\rho_m$ 
to be proportional to $\vec e$ (by properly choosing $\zeta$). For $k=1$ it is 
convenient to parametrize $\zeta^\dagger P_m\zeta=(|\vec\rho_m|-\vec\tau\cdot
\vec\rho_m)/(2\pi)$ in terms of constituent locations, $\vec\rho_m=\Delta
\vec y_m\equiv\vec y_m-\vec y_{m-1}$. As in Sect.~\ref{sec:gaugefield} we will 
take $\vec e=\vec\rho_2/|\vec\rho_2|$. 

We obtain a larger class of $\zeta^a$ for which the $\vec\rho_m$ are parallel, 
by taking $\zeta^a$ to be parallel up to a gauge rotation with an element of 
the unbroken subgroup $U(1)^{n-1}\subset SU(n)$ which leaves the holonomy 
unchanged,
\beq
\zeta_a=\rho_a\exp(2\pi i\alpha_a)\zeta,\qquad \alpha_a\equiv\sum_{m=1}^n
\alpha_a^m P_m,\quad\Tr_n\alpha_a=0.
\eeq
This leads to 
\beq
\vec\rho_m^{\,ab}=\rho^a\rho^b\exp(2\pi i(\alpha_b^m-\alpha_a^m))
\Delta\vec y_m,\quad\hat S_m^{ab}=\vec\rho_m^{\,ab}\cdot\Delta
\vec y_m/|\Delta\vec y_m|.\label{eq:Srho}
\eeq
Note that for $Sp(1)$ we verify that $\vec\rho_1+\vec\rho_2^{\,t}=\vec 0$ and 
$\hat S_1-\hat S_2^t=0$ (using $\alpha_a^1=-\alpha_a^2$). With $\Delta\vec y_m
=\Delta y_m\vec e$ for all $m$ (by definition $\Delta y_2>0$) we may solve 
\refeq{jumps},
\beq
Y_m^{ab}=(\xi_a+\rho_a^2 y_m)\delta_{ab}+i(1-\delta_{ab})\rho_a\rho_b
\sum_{j= 1}^n\Delta y_j\frac{\exp\left(2\pi i\left[\alpha_b^j-\alpha_a^j
-(\mu_j+s_j^m)(\xi_0^b-\xi_0^a)\right]\right)}{2\sin\left(\pi
\left[\xi_0^b-\xi_0^a\right]\right)},\label{eq:Ymfull}
\eeq
where the $\xi_a$ are arbitrary, $m=1,\ldots,n$ and $s_j^m=\half$ for $j=1,2,
\ldots,m$ and $s_j^m=-\half$ for $j=m+1,\ldots, n$. The eigenvalues of these 
hermitian matrices determine the constituent locations, all lined-up along 
$\vec e$. It should be noted that there is no reason to expect that all the 
$Y_m$ can be diagonalized simultaneously. We will come back to this in the 
following section. 

\begin{figure}[htb]
\hskip7.4cm $y$ \vskip1.3cm \hskip10.5cm $\xi$ \vskip1.7cm
\includegraphics{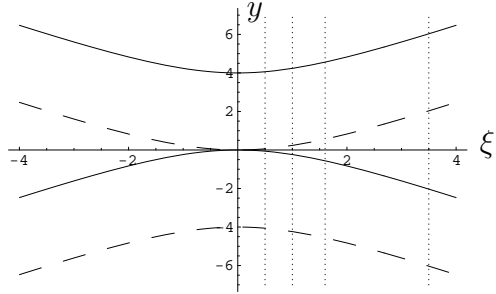}
\caption{Constituent locations $y_m^{(j)}$ based on \protect\refeq{Ym} 
(i.e. $\alpha_a=0$ and $\xi_0\to0$) as a function of $\xi=\xi_1=-\xi_2$ 
for $y_2= -y_1=\nu_1=\nu_2=\half$ and $\rho_1=\rho_2=2$. Dashed versus 
full lines distinguish the magnetic charge of the constituents. The 
dotted lines represent the four cases shown in Figs.~\ref{fig:Wellsep} 
and \ref{fig:Overlap-xz}.}\label{fig:Ydiag}
\end{figure}

Returning to the simplest case of parallel gauge orientations, i.e. putting
$\alpha^j_a=0$, we may take the limit $\xi_0\rightarrow0$ (related to 
vanishing time separations) to find
\beq
Y_m^{ab}=(\xi_a+y_c\rho_a^2)\delta_{ab}+(y_m-y_c)\rho_a\rho_b\quad 
y_c\equiv\sum_{m=1}^n\nu_my_m.\label{eq:Ym}
\eeq 
The ($k=1$) ``center of mass" coordinate $y_c$ can be freely chosen and the
$\xi^a$ play the role of ``center of mass" of each constituent caloron. How
exactly this is realized becomes clear when we diagonalize $Y_m$. Let us
first consider \refeq{Ym} for $SU(2)$ and charge 2, $k=n=2$, with $\xi_1=-
\xi_2\equiv\xi$. Without loss of generality we choose $y_c=0$, such that
the two eigenvalues of $Y_m$ are given by $y_m^{(j)}=\half y_m\rho^2+(-1)^j
\sqrt{\xi^2+\quarter y_m^2\rho^4+y_m\xi\Delta\rho^2}$, where $\rho^2\equiv
\rho_1^2+\rho_2^2$ and $\Delta\rho^2\equiv\rho_1^2-\rho_2^2$. For large and 
positive $\xi$ we find $y_m^{(j)}=(-1)^j\xi+y_m\rho_j^2+\cO(\xi^{-1})$, 
representing two charge 1 calorons centered at $\xi$ and $-\xi$, with 
separations between their constituents monopoles given in terms of $\Delta 
y_2\rho_{1,2}^2$. We plot the constituent locations as a function of $\xi$ 
in Fig.~\ref{fig:Ydiag} for $y_2=-y_1=\nu_j=\half$ and $\rho_j=2$. 

\begin{figure}[htb]
\vskip3cm
\includegraphics{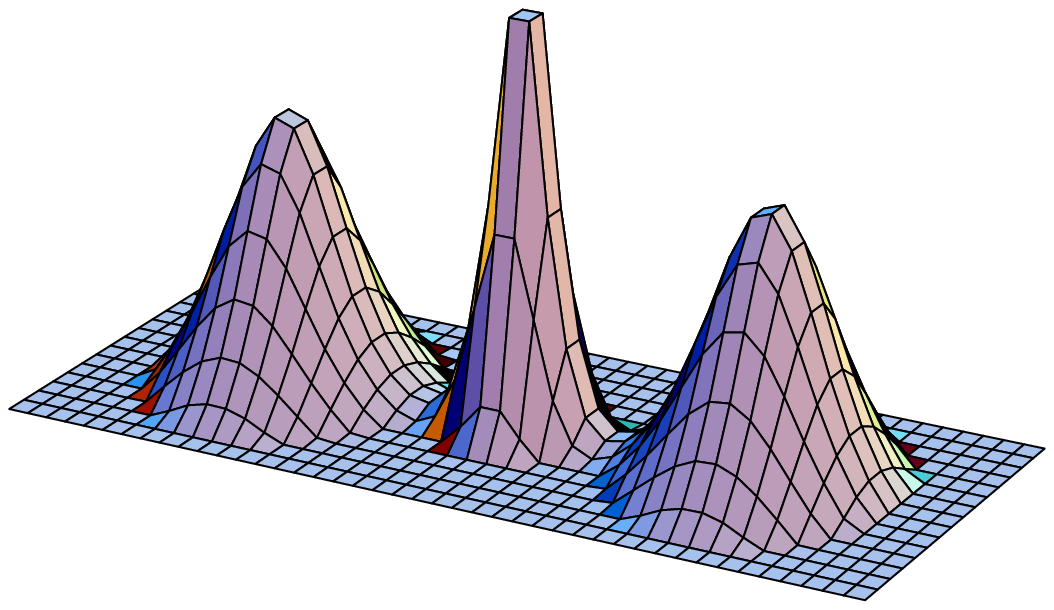}   
\includegraphics{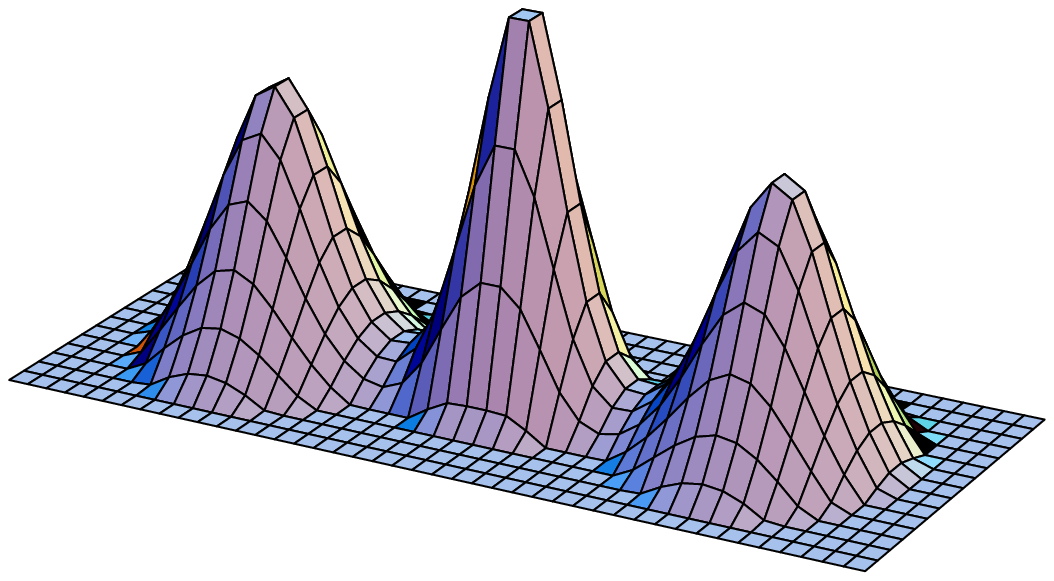}   
\includegraphics{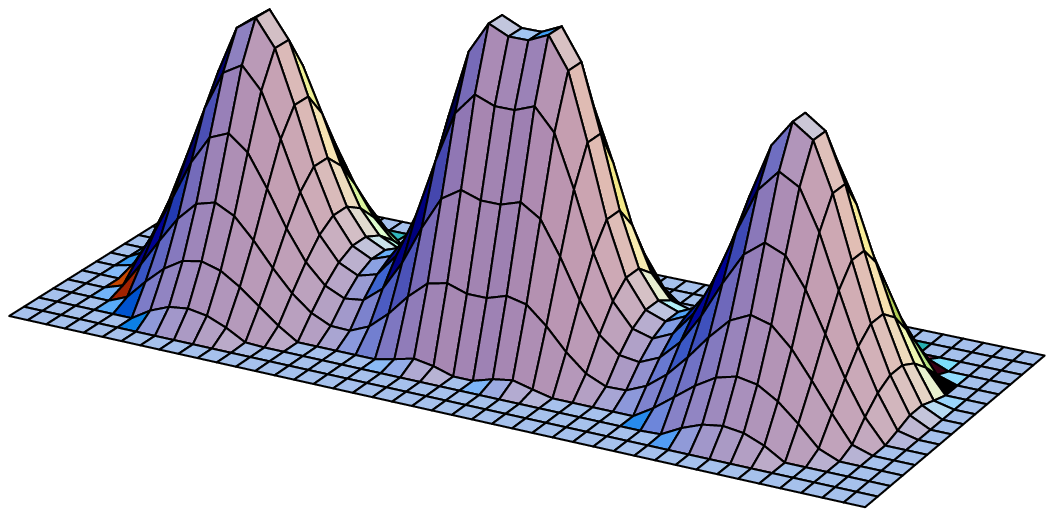}  
\caption{The action density (cutoff for $\log(S)$ below -3) as a function of 
$x$ and $z$ for the $SU(2)$ solution with charge 2 ($\mu_2=\quarter,\,
\alpha_a^j=\xi^a_0=0,\,\rho_j=2$) and increasing values of $\xi\equiv\xi_1=
-\xi_2$ (see Fig~\ref{fig:Wellsep} (left) for $\xi=3.5$), with $\xi=1.6$ 
(left), $\xi= 1.0$ (middle) and $\xi=0.5$ (right). Compare Fig.~\ref{fig:Ydiag} 
for the corresponding constituent locations.}\label{fig:Overlap-xz}
\end{figure}

\begin{figure}[htb]
\vskip6cm
\includegraphics{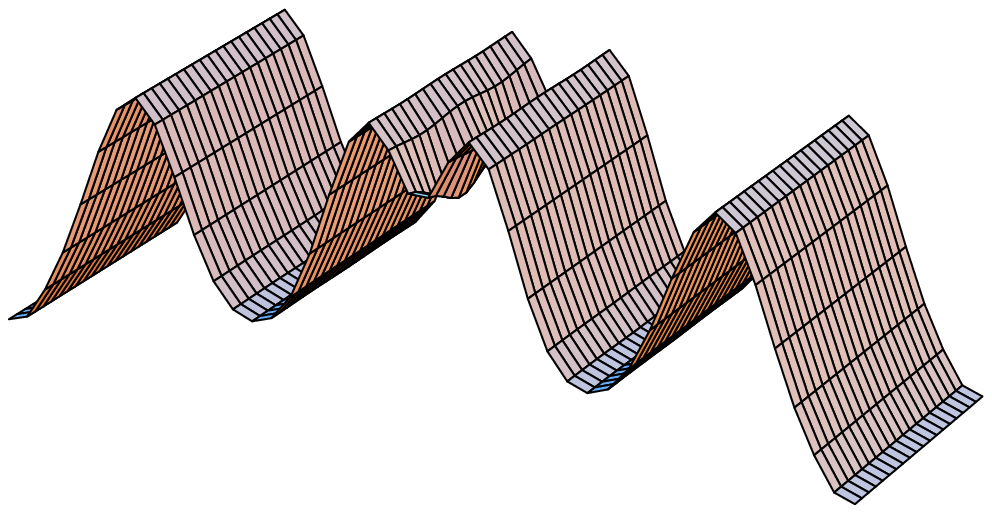}   
\includegraphics{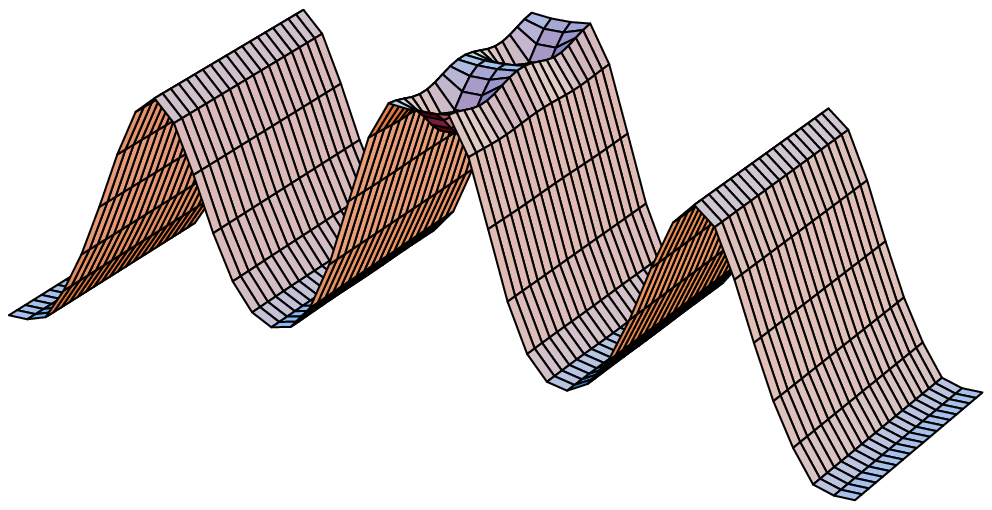}   
\includegraphics{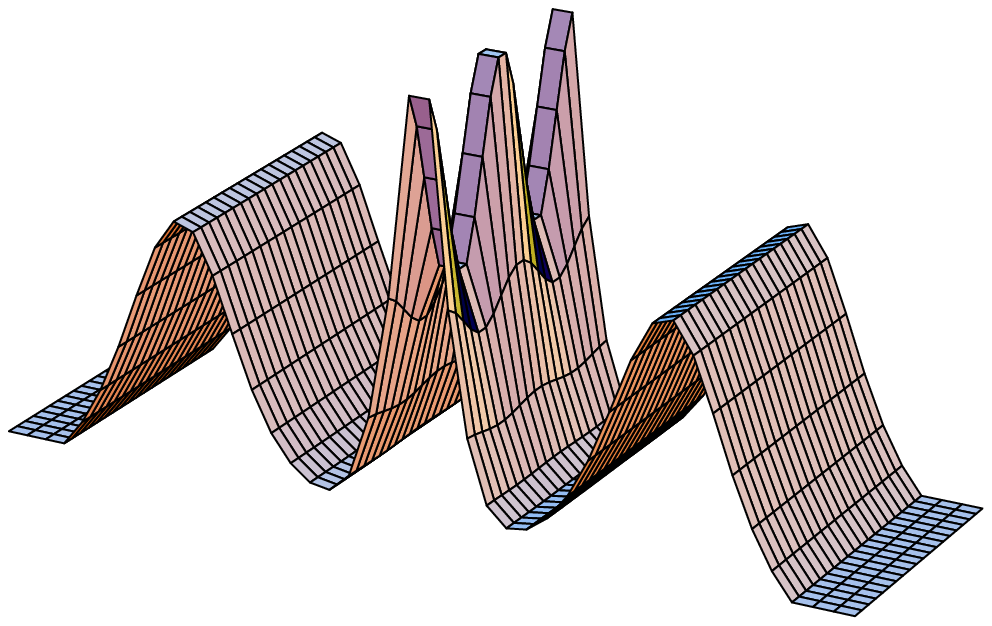}  
\includegraphics{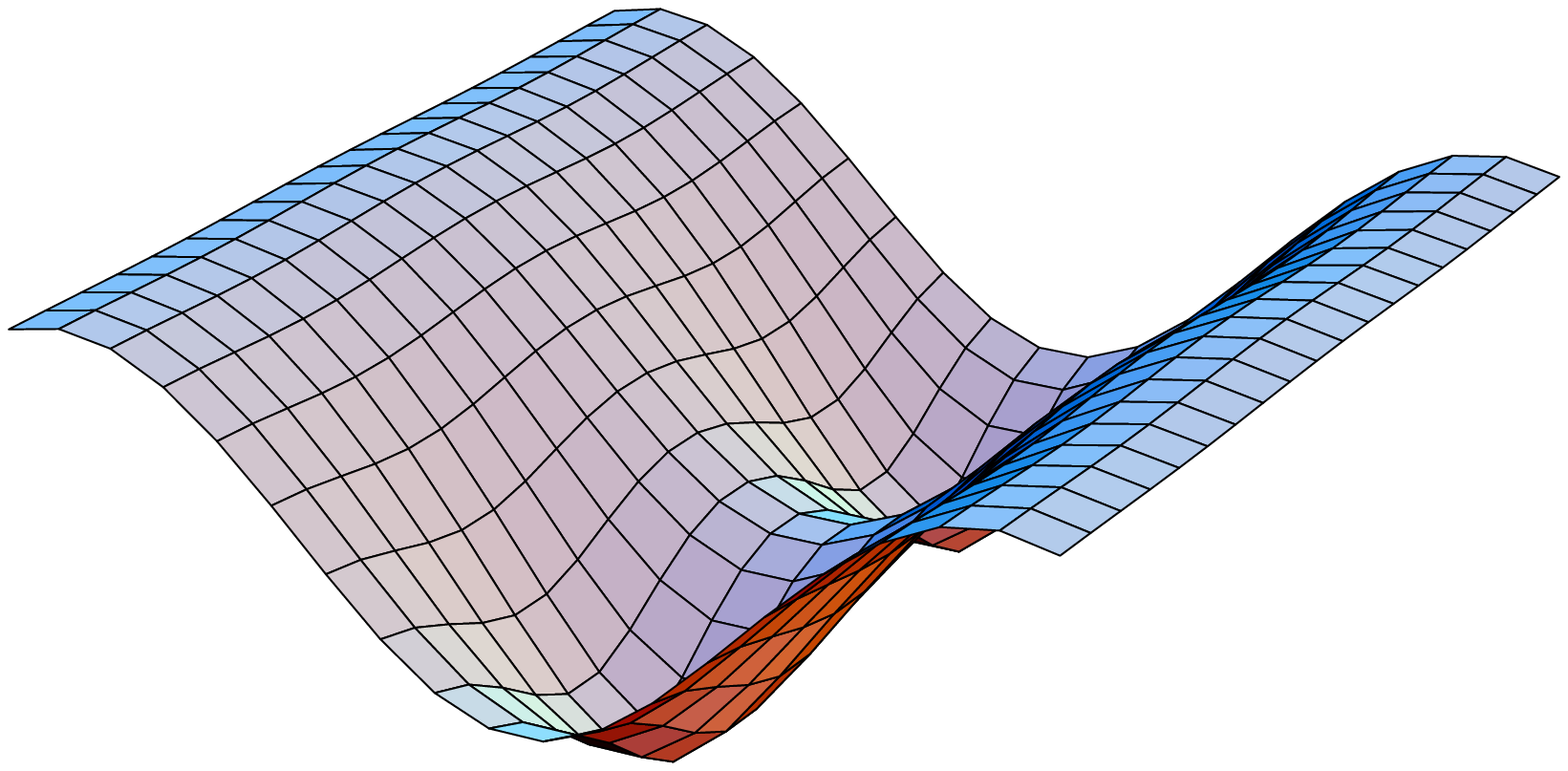}   
\includegraphics{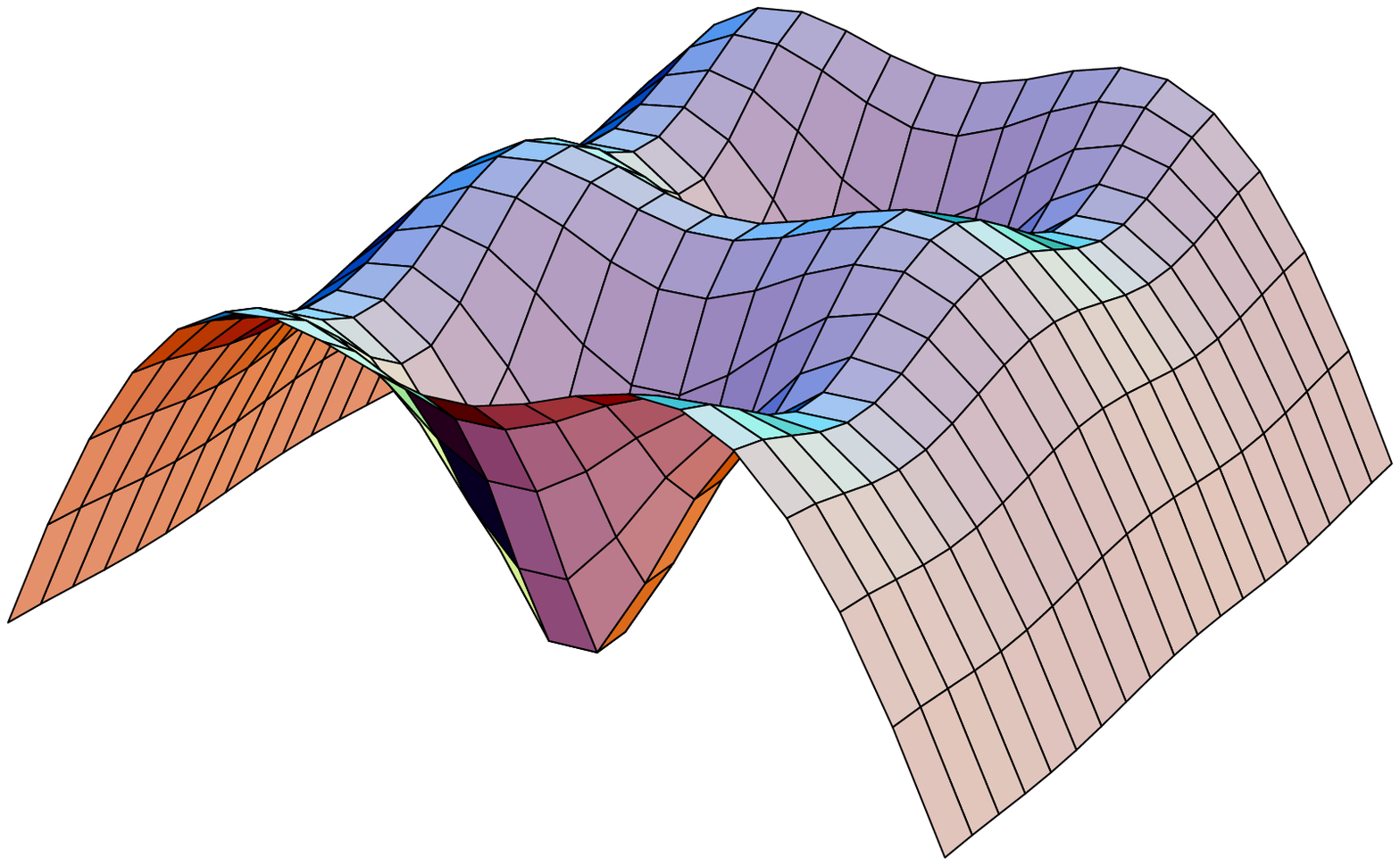}   
\includegraphics{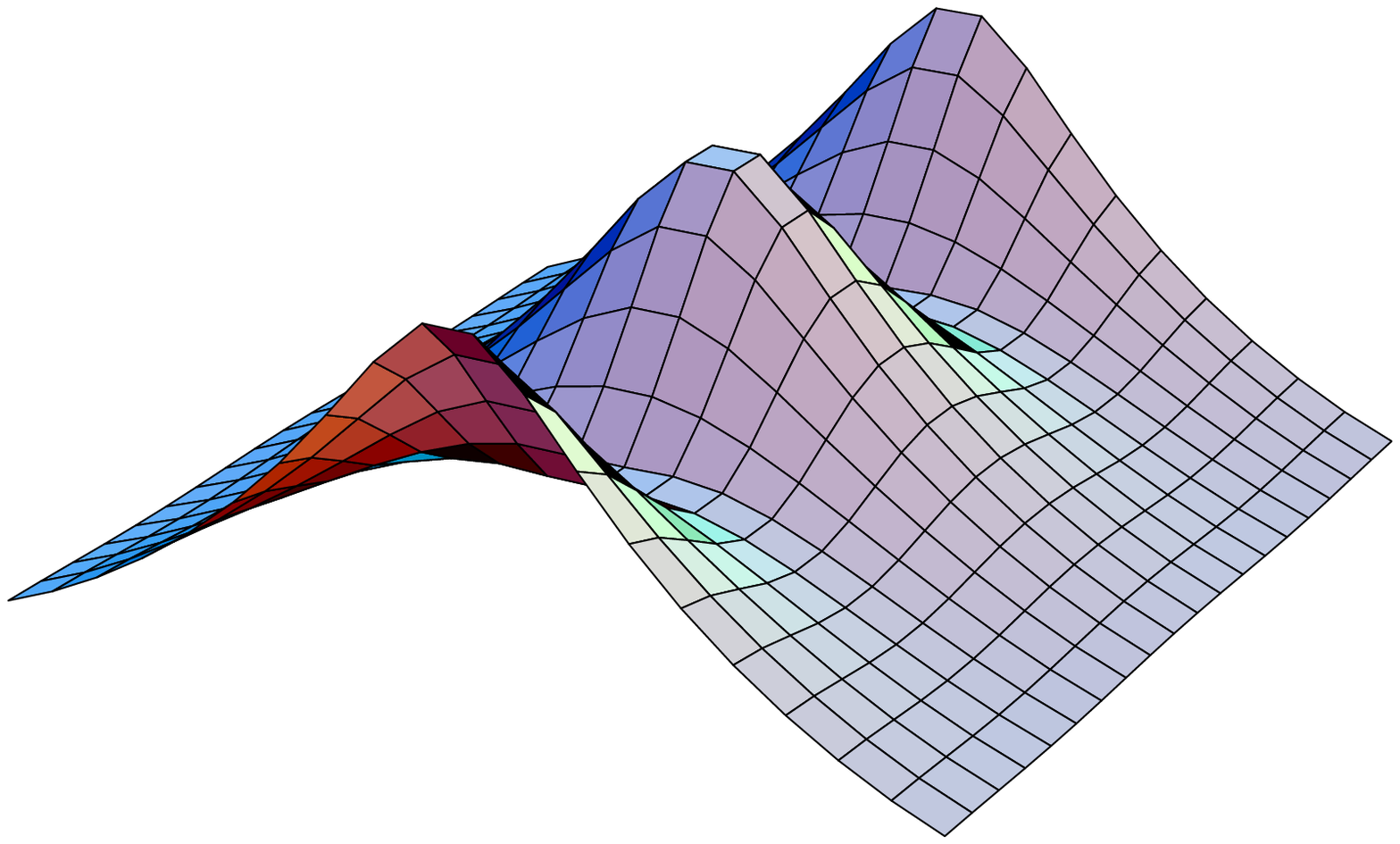}  
\caption{The action density (cutoff for $\log(S)$ below -3) as a function of 
$z$ and $t$ (doubling the time-period) for the $SU(2)$ solution with charge 2 
($\mu_2=\quarter,\,\xi=\half,\,\alpha_a^j=0,\,\rho_j=2$) and increasing values 
of $\xi_0$, with $\xi_0=0$ (left), $\xi_0=0.2$ (middle) and $\xi_0=0.25$
(right), for the top row all on the same scale, zooming in on the middle region
on the bottom row (not to scale). See Fig~\ref{fig:Overlap-xz} (right) for 
the case of $\xi_0=0$ shown as a function of $x$ and $z$.}
\label{fig:Overlap-zt}
\end{figure}

Action density profiles are shown in Fig.~\ref{fig:Wellsep} (left) for $\xi
=3.5$ and in Fig.~\ref{fig:Overlap-xz} for $\xi=1.6,\,1.0,\,0.5$. From the 
dotted lines in Fig.~\ref{fig:Ydiag} one reads off the associated constituent 
locations. Note that the magnetic moments of the two calorons are pointing in 
the same direction and that we can not freely interchange constituent monopole 
locations within our axially symmetric ansatz. However, when $\xi$ is small 
it is more natural to interpret the configuration as a narrow caloron (i.e. 
instanton) with inverted magnetic moment in the background of a large caloron. 
This is the proper setting to understand the non-trivial time dependence for 
$\xi=0.5$ illustrated in Fig.~\ref{fig:Overlap-zt} (left). 

For $\xi\rightarrow0$ a singular caloron arises due to the fusion of two 
constituents (with opposite magnetic charge). This singularity is avoided
when $\xi^a_0\neq 0$, which can be understood by observing that the eigenvalues
of $\xi^a_0$ parametrize time-locations. If $\alpha^j_a\neq 0$, with $\xi_0$ 
and $\xi$ made small, one will find two calorons (and their constituents) to 
be pushed far from each other. This can be understood as well, in terms of a 
short-to-long distance duality in the ADHM data for an instanton pair with 
non-parallel group orientation~\cite{GaKo}, but can also be read off from 
the eigenvalues of $Y_m$ defined in \refeq{Ymfull}. As an example we take 
again $SU(2)$ and charge 2, but now with $\xi_0\equiv\xi_0^1=-\xi_0^2$ and 
$\alpha_1^2=-\alpha_2^2=-\alpha_1^1=\alpha_2^1\equiv\alpha$ in general 
non-zero. 
For the case $\alpha=1/8$ (perpendicular relative color orientations), $\rho_1
=\rho_2\equiv\rho$ and $\mu_2=\quarter$ (equal mass constituents), the 
eigenvalues are  $y_m^{(j)}=y_m\rho^2+(-1)^j\sqrt{\xi^2+\quarter(\Delta y_2)^2
\rho^4\sin^{-2}(\pi\xi_0)}$, whereas for $\alpha=0$ one finds $y_m^{(j)}=y_m
\rho^2+(-1)^j\sqrt{\xi^2+\quarter(\Delta y_2)^2\rho^4\cos^{-2}(\pi\xi_0)}$. 
In the light of this it is interesting to observe, as shown in 
Fig.~\ref{fig:Overlap-zt}, that with $\alpha=0$ and {\em increasing} $\xi_0$ 
the constituents are pushed out in the $z$ direction as well. When $\xi_0\to
0.5$ the constituents would otherwise come close together through the 
periodicity in the time direction.  Effectively these constituents thus have 
perpendicular color orientations (due to our choice of holonomy with $\mu_2=
\quarter$). The transition from constituents separating in the time direction 
for $\xi_0$ near 0 to constituents separating in the $z$ direction for $\xi_0$ 
near $\half$ occurs for $\rho=2$ at approximately $\xi_0=0.2$. 

With a little imagination one detects the ring-shaped structure also 
observed~\cite{GaKo} in the case of instantons at zero temperature, see
Fig.~\ref{fig:Overlap-zt} (middle). A more direct analogy of course occurs
when two calorons (with $\rho$ small, i.e. instantons with unresolved 
constituent monopoles) approach each other. We checked that for $\xi=
\alpha=0$ and $\xi_0\to0$ a singular caloron forms due to the overlap of two 
calorons with parallel gauge orientations, whereas for $\xi_0\to\half$ the 
two calorons are pushed away (to infinity) in the $z$ direction as is 
appropriate for the non-parallel group orientation due to the non-trivial 
holonomy. At an intermediate value ($\xi_0=0.25$ for $\rho_1=\rho_2=0.1$) 
one observes a small ring in the $t$-$z$ plane. Choosing $\xi$ large one 
may check that $\pm\xi_0$ indeed gives the time location for each caloron. 
When, however, $\xi_0$ approaches $\half$ they can no longer keep parallel 
gauge orientations due to the non-trivial holonomy. As noted before, this 
may be described by a solution with $\alpha\neq0$ and $\xi_0\to0$. Computing 
the eigenvalues of $Y_m$ therefore allows one to easily predict the behavior 
of the exact solution. 

For charge 1 it had been shown~\cite{KvBn} that as soon as one of the 
constituents is far removed from the others the solution becomes static. For 
the ``dimensional reduction" to take place at higher charge this is no longer 
sufficient. We have seen (generalization to $SU(n)$ is straightforward) that 
any magnetically neutral cluster of constituents, when small with respect to 
$\beta$, will behave like an instanton that is localized in time. For the 
special case with parallel group orientations, putting all $\xi_a =0$ in
\refeq{Ym} one would even be left with $k-1$ singular instantons on top of 
one regular caloron, whose scale parameter is set by $\rho^2=\sum_{a=1}^k
\rho_a^2$, which can be understood from the fact that the matrix $\rho_a
\rho_b$ has rank 1. It does, however, give us one opportunity to go beyond 
the axial symmetry considered so far. When $\xi_a=0$ we could solve the Nahm 
equation for parallel gauge orientations by $\vec Y^{ab}_m=\vec y_m\rho_a
\rho_b$ {\em without} insisting all the $\vec y_m$ line-up. This still 
describes $k-1$ singular instantons on top of one regular caloron, except 
that now the singular instantons can have {\em arbitrary} locations.

Even though our ansatz to obtain exact solutions has been restrictive 
(as is clear from the axial symmetry), we stress that the solutions for 
the Nahm equation we found provide genuine multi-caloron solutions, which 
reveal $kn$ isolated lumps for each of its constituent monopoles (with 
suitably chosen $\xi^a$ so the constituents do not overlap). This is not 
only illustrated in Fig.~\ref{fig:Wellsep}, but can also be understood 
analytically for any charge $k$ and $SU(n)$ as follows. We diagonalize each 
$Y_m$ with (in general {\em different}) similarity transformations $U_m$, 
or $Y_m\equiv U_m\diag(y_m^{(1)},\ldots,y_m^{(k)})U_m^\dagger$. These bring
$R_m$ to the diagonal form $R_m^\diag=\diag(r_m^{(1)},\ldots,r_m^{(k)})$, 
with $r_m^{(j)}\equiv|\vec x-y_m^{(j)}\vec e|$ the constituent radii, such
that $\tilde H_m\equiv U_m^\dagger H_m U_m$ ($U_m$ acting componentwise) 
simplifies to (cf. \refeq{conHm})
\beq
\tilde H_m\equiv\pmatrix{\cosh(2\pi\nu_m R^\diag_m)&(2\pi R^\diag_m)^{-1}
\sinh(2\pi\nu_m R^\diag_m)\cr2\pi R^\diag_m\sinh(2\pi\nu_m R^\diag_m)&
\cosh(2\pi\nu_m R^\diag_m)\cr}.
\eeq
The action density can now be explicitly expressed in terms of the constituent 
radii
\beqa
&-\half\Tr_n F_{\mu\nu}^2(x)=-\half\partial_\mu^2\partial_\nu^2\log 
\psi(x),\quad\psi=\det(ie^{-\pi ix_0}(\ein_{2k}-\cF)/\sqrt{2}),\nonumber\\
&\cF\equiv\exp(2\pi i(x_0\ein_k-\xi_0))U_n\tilde H_n\tilde T_n\tilde H_{n-1}
\tilde T_{n-1}\cdots\tilde H_1 U_1^\dagger T_1.
\eeqa
cf. Eqs.~(\ref{eq:THs},\ref{eq:Osborn},\ref{eq:defpsi},\ref{eq:conHm}), where 
$\tilde T_m\equiv U_m^\dagger T_mU_{m-1}$ ($U_m$ again acting componentwise).
The size of the constituent monopoles is read off to be $(2\pi\nu_m)^{-1}$ 
(or $\beta(2\pi\nu_m)^{-1}$ when $\beta\neq1$), and one concludes that the 
action density will contain $kn$ lumps for sufficiently well separated 
constituents. The figures were produced by computing the action density
using precisely this method.

\section{Far-field limit}

The non-trivial value of the Polyakov loop at spatial infinity (holonomy) 
leads to a spontaneous breaking of the gauge symmetry, but without the 
need of introducing a Higgs field. One may view $A_0$ as the Higgs field
in the adjoint representation. This is one way to understand why constituent
monopoles emerge. The best way to describe the caloron solutions, in case of
well separated constituents, is by analyzing the field outside the cores of
these constituents, where only the abelian field survives. Since in our case
the asymptotic Polyakov loop value defines a global direction in color space
the abelian generator in terms of which we can describe the so-called far-field 
configuration is fixed, giving rise to a global embedding in the full gauge 
group. Extrapolating the abelian fields back to inside the core of the 
constituents leads to Dirac monopoles. Such an extrapolation is well defined 
in terms of the {\em high temperature limit}, which makes the core of the 
constituents shrink to zero size and the field to become a smooth abelian 
gauge field everywhere except for the singularities of the Dirac monopoles. 
Thus we anticipate that in this limit the self-dual abelian field is still 
described by point like constituents, despite the fact that $\vec R(z,\vec x)$ 
is no longer piecewise constant. Any ``fuzziness" of the constituent location 
that may result from this, would be confined to the non-abelian core, and 
not visible from afar.

\subsection{Green's function}
Despite the somewhat formal expression for the Green's function 
$\hat f_x(z,z')$, one can extract information about the long-range fields 
from it. In the following we will show how to neglect the exponentially
decaying fields in the cores of the monopole constituents, being left
with the abelian components of fields which decay algebraically. We only 
need to consider the ``bulk" contributions $H_m$, \refeq{Hm}, which contain
all the dependence on $\vec x$. Our starting point is \refeq{dftrans} 
restricted to the $m$th interval, $z\in(\mu_m,\mu_{m+1})$
\beq
\left\{-\frac{d^2}{dz^2}+4\pi^2\vec R^2(z;\vec x)\right\}f_m(z)=0.
\label{eq:dfm}
\eeq
To distinguish between exponentially growing and decreasing contributions for
this homogeneous equation we take as a basis for $f_m(z)$ functions 
$f_m^\pm(z)$ with the following asymptotic behavior
\beq
|\vec x|\to\infty:\:f_m^\pm(z)\to\exp\left(\pm 2\pi|\vec x|(z-\mu_m)
\ein_k\right),\label{eq:flim}
\eeq
relying on the fact that $\vec R^2(z;\vec x)\to\vec x^{\,2}\ein_k$. This
prompts us to introduce on each interval the matrix valued functions 
$R^\pm_m(z)$ (we suppress the dependence on $\vec x$) such that
\beq 
f_m^\pm(z)=P\exp\left[\pm 2\pi\int_{\mu_m}^z \!\!\!R_m^\pm(z)dz\right].
\eeq
from which it follows that $R^\pm_m(z)$ is a solution of the Riccati equation
\beq
R_m^\pm(z)^2\pm\frac{1}{2\pi}\frac{d}{dz}\,R_m^\pm(z)=\vec
R^2(z;\vec x).\label{eq:Riccati} 
\eeq
We note that for $|\vec x|\to\infty$, $R_m^\pm(z)\to|\vec x|$ and that for 
piecewise constant $\vec R(z;\vec x)$ both $R^+_m(z)$ and $R^-_m(z)$ are 
constant and equal to $R_m$, introduced in \refeq{conHm}. 

We can write for $z,z'\in(\mu_m,\mu_{m+1})$ the ``propagator" $W(z,z')$
defined in \refeq{Wdef} in terms of $f_m^\pm(z)$ as $W(z,z')=W_m(z)
W^{-1}_m(z')$ with
\beq
W_m(z)\equiv\pmatrix{f_m^+(z)&f_m^-(z)\cr 2\pi R^+_m(z)f_m^+(z)&-2\pi 
R^-_m(z)f_m^-(z)\cr}.
\eeq
Using that $H_m=W_m(\mu_{m+1})W_m^{-1}(\mu_m)$ and $f^\pm_m(\mu_m)=\ein_k$, 
we find by neglecting the exponentially decreasing factors $f_m^-(\mu_{m+1})$ 
the required limiting behavior for $H_m$. 
Paying special attention to the ordering of the $k\times k$ matrices $R_m^\pm$,
observing that
\beq
W_m^{-1}(\mu_m)=(4\pi R_m(\mu_m))^{-1}\pmatrix{2\pi R^-_m(\mu_m)&\ein_k\cr2\pi 
R^+_m(\mu_m)&-\ein_k\cr},\quad R_m\equiv\half(R_m^+(\mu_m)+R_m^-(\mu_m)),
\eeq
we find well outside the cores of the constituents
\beq
H_m\to\pmatrix{\ein_k&0\cr 2\pi R_m^+(\mu_{m+1})&0\cr}f_m^+(\mu_{m+1})
(4\pi R_m)^{-1}\pmatrix{2\pi R_m^-(\mu_m)&\ein_k\cr0&0\cr}.
\eeq
The sparse nature of the matrices involved will be of considerable help to 
simplify the limiting behavior of the Green's function. A crucial ingredient 
is the combination
\beqa
&&\pmatrix{2\pi R_m^-(\mu_m)&\ein_k\cr 0&0\cr}\pmatrix{\ein_k&0\cr 2\pi S_m&
\ein_k\cr}\pmatrix{\ein_k&0\cr 2\pi R_{m-1}^+(\mu_m)&0\cr}=\pmatrix{2\pi
\Sigma_m&0\cr 0&0\cr}\nonumber\\
&&\Sigma_m\equiv R_{m-1}^+(\mu_{m})+R_m^-(\mu_m)+S_m,\label{eq:defSigma}
\eeqa
for clarity summarizing the various ingredients in the following picture\\
\setlength{\unitlength}{0.91cm}
\thinlines
\begin{picture}(10,3.6)(-0.5,-1.2)
\put (1.2,0){\line(1,0){0.2}}
\put (1.6,0){\line(1,0){0.2}}
\put (0.95,-0.5){$z=$}
\put (2,0){\line(1,0){11}}
\put (2,-0.25){\line(0,1){0.5}}
\put (2,0){\circle*{0.1}}
\put (1.7,-0.5){$\mu_{m-1}$}
\put (1.7,1.3){${}_{\Sigma_{m-1}}$}
\put (1.7,0.7){${}_{R_{m-1}}$}
\put (3.2,-0.6){$R^\pm_{m-1}(z),f^\pm_{m-1}(z)$}
\put (7.75,-0.25){\line(0,1){0.5}}
\put (7.75,0){\circle*{0.1}}
\put (7.45,-0.5){$\mu_m$}
\put (7.45,1.7){${}_{\Sigma_m=}$}
\put (5.7,1.3){${}_{R^-_{m-1}(\mu_m)+S_m+R^+_m(\mu_m)}$}
\put (7.45,0.8){${}_{R_m=}$}
\put (7.75,0.5){${}_{(R^+_m(\mu_m)+R^-_m(\mu_m))/2}$}
\put (9.1,-0.6){$R^\pm_m(z),f^\pm_m(z)$}
\put (13,-0.25){\line(0,1){0.5}}
\put (13,0){\circle*{0.1}}
\put (12.7,-0.5){$\mu_{m+1}$}
\put (12.7,1.3){${}_{\Sigma_{m+1}}$}
\put (12.7,0.7){${}_{R_{m+1}}$}
\put (13.2,0){\line(1,0){0.2}}
\put (13.6,0){\line(1,0){0.2}}
\put (14.4,0.){.}
\end{picture}
\setlength{\unitlength}{1.0pt}\\
This leads to the following far-field approximation for $\cF_{\mu_m}$ 
(cf. \refeq{THs}),
\beq
\cF_{\mu_m}\to\hat g^\dagger(1)\pmatrix{\ein_k&0\cr 2\pi(R_{m+n-1}^+(\mu_{m+n})
+S_{m+n})&0\cr}\cG_m\pmatrix{2\pi R_m^-(\mu_m)&\ein_k\cr 0&0\cr}
\label{eq:approxF}
\eeq
where $\cG_m\equiv\cG_{m+n,m}$ and
\beqa
\cG_{m',m}&\equiv&f_{m'-1}^+(\mu_{m'})(2R_{m'-1})^{-1}\Sigma_{m'-1}
        f_{m'-2}^+(\mu_{m'-1})(2R_{m'-2})^{-1}\Sigma_{m'-2}\nonumber\\
&&\cdots\cdots f_{m+1}^+(\mu_{m+2})(2R_{m+1})^{-1}\Sigma_{m+1}
               f_m^+(\mu_{m+1})(4\pi R_m)^{-1}.\label{eq:defGm}
\eeqa
One might have expected a factor $2\pi\Sigma_m$ on the right, but this is 
contained in the remaining terms of \refeq{approxF}. For example $\Tr_k(
\cF_{\mu_m})=\Tr_k\left(\hat g^\dagger(1)\cG_m(2\pi\Sigma_m)\right)$.

As we have seen in Eqs.~(\ref{eq:Adef}) and (\ref{eq:phidef}), the gauge field 
only requires us to know the Green's function at the impurities. Without loss 
of generality we may assume $\mu_{m'}>\mu_m$ and take $z_0=\mu_m+0$, such that 
(see Eqs.~(\ref{eq:ftofh},\ref{eq:vecf},\ref{eq:finf}))
\beq
f_x(\mu_{m'},\mu_m)=-4\pi^2\left(\ein_k,0\right)\cdot W(\mu_{m'},\mu_m+0)
(\ein_{2k}-\cF_{\mu_m})^{-1}\pmatrix{0\cr\ein_k}.
\eeq
The matrix $(\ein_{2k}-\cF_{\mu_m})$ has a $2\times 2$ block structure, and 
one can verify that in general
\beq
\pmatrix{a&b\cr c&d\cr}^{-1}=\pmatrix{(a-bd^{-1}c)^{-1}&(c-db^{-1}a)^{-1}\cr
(b-ac^{-1}d)^{-1}&(d-ca^{-1}b)^{-1}\cr}.\label{eq:blockinv}
\eeq
Identifying the blocks, in the high temperature limit we find
\beqa
a\equiv(\ein_{2k}-\cF_{\mu_m})_{11}&\to&\ein_k-2\pi\hat g^\dagger(1)\cG_m 
R_m^-(\mu_m),\nonumber\\
b\equiv(\ein_{2k}-\cF_{\mu_m})_{12}&\to&-\hat g^\dagger(1)\cG_m,\nonumber\\
c\equiv(\ein_{2k}-\cF_{\mu_m})_{21}&\to&-4\pi^2(R_{m-1}^+(\mu_m)+
S_m)\hat g^\dagger(1)\cG_m R_m^-(\mu_m),\label{eq:highT}\\
d\equiv(\ein_{2k}-\cF_{\mu_m})_{22}&\to&\ein_k-2\pi(R_{m-1}^+(\mu_m)+S_m)
\hat g^\dagger(1)\cG_m\nonumber,
\eeqa
where we used that $R_{m+n-1}^\pm(\mu_{m+n})+S_{m+n}\!=\!\hat g(1)(R_{m-
1}^\pm(\mu_m)+S_m)\hat g^\dagger(1)$, cf. Eqs.~(\ref{eq:Frel},\ref{eq:THs}). 
Evaluating the Green's function at the {\em same} impurities, $\mu_{m'}=
\mu_m$, is simplified by the fact that $W(\mu_m,\mu_m)=\ein_{2k}$. This
gives the following remarkably simple result in the far-field limit,
\beq
f_x(\mu_m,\mu_m)=-4\pi^2(\ein_{2k}-\cF_{\mu_m})_{12}^{-1}\to4\pi^2
\left(2\pi\Sigma_m-\cG_m^{-1}\hat g(1)\right)^{-1}\!\!\!\to 
2\pi\left(\Sigma_m\right)^{-1}.\label{eq:fdiag}
\eeq

For the Green's function evaluated at {\em different} impurities, $\mu_m\neq
\mu_{m'}$, we need to determine $W(\mu_{m'}-0,\mu_m+0)$, for which we can
follow the same method as for $\cF_{\mu_m}$ 
\beq
W(\mu_{m'}-0,\mu_m+0)=\pmatrix{\ein_k&0\cr 2\pi R_{m'-1}^+(\mu_{m'})
&0\cr}\cG_{m',m}\pmatrix{2\pi R_m^-(\mu_m)&\ein_k\cr 0&0\cr},
\eeq
with $\cG_{m',m}$ as defined in \refeq{defGm}. This leads to 
\beqa
f_x(\mu_{m'},\mu_m)&\to&-4\pi^2\cG_{m',m}\left((\ein_{2k}-\cF_{\mu_m}
)^{-1}_{22}+2\pi R_m^-(\mu_m)(\ein_{2k}-\cF_{\mu_m})^{-1}_{12}\right)
\nonumber\\&\to&\cG_{m',m}\cG_m^{-1}\hat g(1)f_x(\mu_m,\mu_m),
\eeqa
which is exponentially suppressed since $\cG_{m',m}$ grows as $\exp\left(
2\pi|\vec x|(\mu_{m'}-\mu_m)\right)$. This cannot compensate for the decay 
of $\cG_m^{-1}$, provided all $\mu_m$ are unequal, i.e. all constituents 
have a non-zero mass. Massless constituents have a so-called non-abelian 
cloud~\cite{Wein}, which has no abelian far-field limit. 

\subsection{Total action}

To determine $\psi$ in the expression for the action density, 
Eqs.~(\ref{eq:Osborn},\ref{eq:defpsi}), we need to compute 
$\det(\ein_{2k}-\cF_{\mu_m})$. Using \refeq{highT} we find
\beqa
&\det\left(\ein_{2k}-F_{\mu_n}\right)=\det\pmatrix{a&b\cr c&d\cr}=\det\pmatrix{
0&b\cr c-ab^{-1}d&d\cr}=\det(b)\det(ab^{-1}d-c)\nonumber\\&\hskip1cm\to\det
\left(\hat g^\dagger(1)\cG_m\right)\det\left(\cG_m^{-1}\hat g(1)-2\pi\Sigma_m
\right)\to\det\left(-2\pi\hat g^\dagger(1)\cG_m\Sigma_m\right),
\eeqa
such that
\beq
\psi\to\det(\pi\cG_m\Sigma_m)=2^{-k}\prod_{m=1}^n\left\{\det\left(f^+_m(
\mu_{m+1})\right)\det(\Sigma_m)/\det(2R_m)\right\}.
\eeq
For $|\vec x|\to \infty$, $f_m^+(\mu_{m+1})\to\exp(2\pi\nu_m|\vec x|\ein_k)$ 
(see \refeq{flim}) and $\half\Sigma_m,~R_m\to|\vec x|\ein_k$, which implies 
that $\psi\to2^{-k}\prod_{m=1}^n\det\left[\exp(2\pi\nu_m|\vec x|\ein_k)\right]
=2^{-k}\exp(2\pi k|\vec x|)$ (recall that $\sum_{m=1}^n\nu_m=1$). Therefore, 
the action is given by $S=-\half\int d^4x~\partial_\mu^2\partial_\nu^2\log
\psi(x)=8\pi^2k$, as should be the case for a self-dual charge $k$ solution.

\subsection{Gauge field}

Without the off-diagonal components of the Green's function contributing 
to the far-field region, the functions $\phi$ and $\phi_j$ in 
Eqs~(\ref{eq:Adef},\ref{eq:phidef}) can be further simplified to
\beq
\phi(x)^{-1}\to1-\sum_m\hat f^{ab}(\mu_m,\mu_m)P_m\zeta_a\zeta^\dagger_b P_m,
\qquad\phi_j\to\sum_{m}\hat f^{ab}(\mu_m,\mu_m)P_m\zeta_a \hat\sigma_j
\zeta^{\dagger}_bP_m,
\eeq
and only the abelian components of the gauge field survive. Particularly the 
case of $Sp(1)$ discussed in Sect.~\ref{sec:gaugefield} is easy to deal with. 
Using 
Eqs.~(\ref{eq:ftofh},\ref{eq:ASp1},\ref{eq:phis},\ref{eq:vphis},\ref{eq:fdiag}) 
we find 
\beqa
\phi^{-1}(x)&\!\!\to\!\!&\sigma_0(1-\Tr_k[2\Sigma_2^{-1}S_2])\equiv\sigma_0
\phi_{\rm ff}^{-1}(x),\quad\vec\phi(x)\to\hat\omega\cdot\vec\sigma\,
\Tr_k[2\Sigma_2^{-1}\hat g(\mu_2)\vec\rho_2\hat g^\dagger(\mu_2)]),
\nonumber\\A_\mu(x)&\!\!\to\!\!&\frac{i}{2}\hat\omega\cdot\vec\tau(1-\Tr_k[2
\Sigma_2^{-1}S_2])^{-1}\bar\eta_{\mu\nu}^j\partial_\nu\Tr_k[2\Sigma_2^{-1}
\hat g(\mu_2)\rho_2^j\hat g^\dagger(\mu_2)]),\label{eq:approxphi}
\eeqa
where we recall (see \refeq{defSigma}) that $\Sigma_2=(R_1^+(\mu_2)+
R_2^-(\mu_2)+S_2)$. This is in perfect agreement with the earlier $k=1$ 
results~\cite{KvB2}. Note that the gauge rotation which relates $\hat 
f_x(z,z')$ to $f_x(z,z')$ also relates $\hat S_m$ to $S_m$ (see 
Eqs.~(\ref{eq:ftofh},\ref{eq:Rdef})) and therefore does not appear 
in the final expression for $\phi(x)$. 

It is interesting to note that the dipole moment of the abelian gauge field
is particularly simple and does not require us to solve for $R_m^\pm(z)$,
since $\lim_{|\vec x|\to\infty}\Sigma_2=2|\vec x|\ein_k$ such that
\beq
\lim_{|\vec x|\to\infty}A_\mu(x)=\frac{i}{2}\hat\omega\cdot\vec\tau\,
\bar\eta_{\mu\nu}^j\partial_\nu\frac{\Tr_k(\rho_2^j)}{|\vec x|}.\label{eq:dip}
\eeq
Hence the dipole moment $\vec p\equiv\half\Tr_k(\vec\rho_2)$ only involves
$\zeta_a$, and we do except it allows for configurations with a vanishing 
dipole moment. For higher multipole moments, through $R_m^\pm(z)$, we need 
to deal with the full quadratic ADHM constraint, or equivalently with the 
Riccati and Nahm equations.  Nevertheless, it is remarkable that in the high 
temperature limit the $\vec x$ dependence is restricted to $R_1^+(\mu_2)$ and 
$R_2^-(\mu_2)$ {\em only}. We would like to prove that each of its eigenvalues
vanish at an isolated point, as one way to identify the $2k$ constituent 
locations. We will defer the study of this interesting issue, and its 
generalization to $SU(n)$, to a future publication.

\subsubsection{Axially symmetric case}

The far-field approximation further simplifies when considering the axially 
symmetric solutions discussed in Sect.~\ref{sec:axial}. We restrict ourselves 
here to $Sp(1)$. Since $\vec R(z;\vec x)$ is piecewise constant, the Riccati 
equation is trivial to solve,
\beq
R^\pm_m(z)=R_m=\sqrt{(\vec x\ein_k-\vec e\,Y_m)\cdot(\vec x\ein_k-\vec e\,Y_m)}.
\eeq
The square root involves a positive $k\times k$ matrix, and is well-defined. 
Due to the fact that $S_m=\hat g(\mu_m)\hat S_m\hat g^\dagger(\mu_m)=\hat g
(\mu_m)\Delta\vec y_m\cdot\vec\rho_m/|\Delta y_m|\hat g^\dagger(\mu_m)$ (see 
Eqs.~(\ref{eq:Rdef},\ref{eq:Srho})), the abelian component of the gauge field 
is of the simple form ($\vec e=\vec\rho_2/|\vec\rho_2|$, see 
Sect.~\ref{sec:gaugefield})
\beq
A^{\rm abel}_\mu(x)=-\frac{i}{2}\hat\omega\cdot\vec\tau e_j\bar\eta^j_{\mu\nu}
\partial_\nu\log\phi(x).
\eeq
In the far field limit $\phi(x)\to\phi_{\rm ff}(x)$ (see \refeq{approxphi}). 
Since $S_2$ has rank 1, the matrix $M\equiv 2\Sigma_2^{-1}S_2$ has only one 
non-vanishing column with respect to a suitably chosen basis, which implies 
that $1-\Tr_k(M)=\det(\ein_k-M)$. This allows us to write in the far-field 
region
\beq
\phi_{\rm ff}(x)=\frac{\det(R_1+R_2+S_2)}{\det(R_1+R_2-S_2)},
\eeq
from which we immediately read off the result~\cite{KvB2} for $k=1$, in which
case it is easy to show that $\phi_{\rm ff}(x)=(r_2+\vec e\cdot\vec r_2)/(r_1
+\vec e\cdot\vec r_1)$, revealing $A_\mu(x)$ to be a linear superposition of 
two oppositely charged self-dual Dirac monopoles. We would like $\phi_{\rm ff}
(x)$ to similarly factorize for $k>1$ in $2k$ Dirac monopoles, but since 
$R_1$, $R_2$ and $S_2$ in general do not commute, some care is required in 
demonstrating the factorization. We will rely on the fact that $(R_m-\vec e
\cdot\vec R_m)(R_m+\vec e\cdot\vec R_m)=(R_m+\vec e\cdot\vec R_m)(R_m-\vec e
\cdot\vec R_m)=R_m^2-(\vec e\cdot\vec R_m)^2=(\vec x\times\vec e)^2\ein_k$ 
and hence independent of $m$. Since by definition $\Delta y_2>0$ (see 
Sect.~\ref{sec:axial}), one finds that
\beq
\phi_{\rm ff}(x)=\frac{\det\left((R_1-\vec e\cdot\vec R_1)+
(R_2+\vec e\cdot\vec R_2)\right)}{\det\left((R_2-\vec e\cdot\vec R_2)+
(R_1+\vec e\cdot\vec R_1)\right)}.
\eeq
This can now be reorganized according to
\beqa
\phi_{\rm ff}(x)&=&\frac{\det\left(\left((R_1-\vec e\cdot\vec R_1)+(R_2+\vec e
\cdot\vec R_2)\right)(R_1+\vec e\cdot\vec R_1)\right)}{\det\left((R_2+\vec e
\cdot\vec R_2)\left((R_2-\vec e\cdot\vec R_2)+(R_1+\vec e\cdot\vec R_1)\right)
\right)}\frac{\det(R_2+\vec e\cdot\vec R_2)}{\det(R_1+\vec e\cdot\vec R_1)}
\nonumber\\&=&\frac{\det(R_2+\vec e\cdot\vec R_2)}{\det(R_1+\vec e\cdot
\vec R_1)},
\eeqa
after which we can separately diagonalize $\vec R_1$ and $\vec R_2$ to find
\beq
\phi_{\rm ff}(x)=\prod_i\frac{r_2^{(i)}+\vec e\cdot\vec r_2^{\,(i)}}{r_1^{(i)}
+\vec e\cdot\vec r_1^{\,(i)}}=\prod_i\frac{r_1^{(i)}+r_2^{(i)}+|y_1^{(i)}-
y_2^{(i)}|}{r_1^{(i)}+r_2^{(i)}-|y_1^{(i)}-y_2^{(i)}|},
\eeq
where $y_m^{(i)}\vec e$ give the locations of the constituent monopoles, with 
$y_m^{(i)}$ the eigenvalues of $Y_m$ and $\vec r_m^{\,(i)}=\vec x-y_m^{(i)}
\vec e$ (the index $m$ distinguishing their charge). The second expression for 
the factorized version of $\phi_{\rm ff}(x)$ uses the fact that the constituent 
locations can be ordered according to
\beq
y_1^{(1)}<y_2^{(1)}<y_1^{(2)}<y_2^{(2)}<\ldots<y_1^{(k)}<y_2^{(k)}.
\eeq
It should be noted that this prevents the constituents to pass each other 
while varying the parameters for the axially symmetric configuration, see 
also Fig.~\ref{fig:Ydiag} and the discussion in Sect.~\ref{sec:axial}. We 
need to go beyond this simple axially symmetric configuration to allow for 
the constituents to rearrange themselves more freely.

\section{Discussion}

We have presented the general formalism for finding exact instanton solutions 
at finite temperature (calorons) with any non-trivial holonomy and topological 
charge. In an infinite volume holonomy and charge are fixed. The solution is 
described by $4kn$ parameters of which $3kn$ give the spatial locations of 
the $kn$ constituent monopoles. The remaining parameters are given by $k$ 
time locations and $(n-1)k$ phases associated to gauge rotations in the 
subgroup that leaves the holonomy unchanged. Of these, $n-1$ can be considered 
as global gauge rotations. The dimension of the moduli space, i.e. the number 
of gauge invariant parameters, is therefore equal to $4kn-(n-1)$. Our subset
of axially symmetric solutions has $2nk+4$ paramters of which there are $n-1$
global gauge rotations, or $2nk-n+5$ gauge invariant paramters.

Explicit solutions were found for the case of axial symmetry, an important
limitation being the difficulty of solving the Nahm equation, or equivalently 
the quadratic ADHM constraint. We certainly expect more progress can be made
on this in the near future. Nevertheless, we already found a rich structure
that bodes well for being able to consider the constituent monopoles as
independent objects. This is surprisingly subtle, as we have illustrated by 
the fact that an approximate superposition of charge 1 calorons tends to give 
rise to a {\em visible} Dirac string. This is also related to the difficulty 
of finding approximate multi-monopole solutions, which can be obtained from 
the caloron solutions by sending a subset of the constituent monopoles to 
infinity, as has been well studied in the charge 1 case~\cite{KvB2,Kraan}. 

An important tool has been our study of the far-field limit, describing the 
abelian gauge field far removed from any of the constituent monopoles. This 
allows for a description of the long distance properties in terms of 
(self-dual) Dirac monopoles. Much could be extracted concerning its properties 
without the need to explicitly solve the Nahm equation. We conjecture in 
general to be able to identify the Dirac monopole location, but some work 
remains to be done here.

\begin{figure}[htb]
\vskip4.6cm
\includegraphics{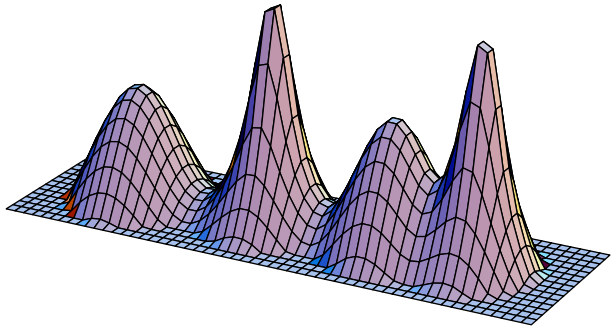}
\caption{The logarithm of the action density (cutoff for $\log(S)$ below -3)
for an $SU(2)$ charge 2 caloron with one type of constituent three times more 
massive than the other ($\mu_2=1/8,\,\alpha_a^j=\xi_0^a=0,\,\rho_j=2,\xi=3.5)$
}\label{fig:uneq}
\end{figure}

It may seem that all these results are somewhat academic since until recently 
none of these constituent monopoles were found in dynamical lattice 
configurations. First of all one would be tempted to search for them at high 
temperature, but it should be noted that above the deconfining phase 
transition the average Polyakov loop takes on trivial values, associated to the 
center of the gauge group, which is not the environment in which a caloron 
will reveal its constituents. This would give the well-known Harrington-Shepard
solution constructed long ago~\cite{HaSh}. With our present understanding this
solution can be seen as having $n-1$ massless constituents which cannot be
localized. Only when sending all these to infinity one is left with a 
monopole~\cite{Ros}. Furthermore, at high temperature classical configurations 
will be heavily suppressed due to their Boltzmann weight. Rather, the hope is 
that the constituent monopoles play an important role {\em below} the 
deconfining temperature, where the average of the Polyakov loop is non-trivial,
and tends to favor {\em equal} mass constituents. This is why in this paper
our examples were for that case, see 
Figs.~\ref{fig:Wellsep},\ref{fig:Overlap-xz},\ref{fig:Overlap-zt} and the 
discussion in Sect.~\ref{sec:axial}. Nevertheless, the formalism developed
here gives results for any choice of the holonomy, and a sample of unequal 
mass constituents is given in Fig.~\ref{fig:uneq}. In general the constituent 
monopoles can be characterized by their magnetic (=electric) charge. For 
$SU(n)$ there are $n$ different types of abelian charges involved~\cite{KvB2}, 
and all $k$ constituents of a given type have the same mass.

We will discuss briefly the lattice evidence for the presence of constituent
monopoles that has accumulated the last few years. A first numerical study 
using cooling was performed with twisted boundary conditions, which implies 
non-trivial holonomy~\cite{MTAP}. Good agreement was found with the infinite 
volume charge 1 analytic results, in particular so for the fermion
zero-modes~\cite{MTCP,MTP} which are more localized than the action density.
A charge 2 solution was also found, shown in Fig.~8 of Ref.~\cite{MTAP}. 
Fermion zero-modes played an intricate role in an extensive numerical study 
of Nahm dualities on the torus~\cite{Tdual}. 

As suggested in Ref.~\cite{KvB2} one may also enforce non-trivial holonomy on 
the lattice by putting at the spatial boundary of the box all links in the 
time direction to the same constant value $U_0$, such that $U_0^{N_t}=\pl$ 
($N_t$ the number of lattice sites in the time direction). This has been 
implemented in $SU(2)$ lattice Monte Carlo studies as well, where $\pl$ was 
set to the average value of the Polyakov loop, appropriate for the temperature 
at which the simulations were performed~\cite{Ilg1,Ilg2}. Cooling was applied 
to find calorons, including those at higher charge. Apart from the 
configurations that in the continuum would be exactly self-dual, the lattice 
allows one to also consider configurations in which both self-dual and 
anti-selfdual lumps appear. This revealed constituent monopoles that seem not 
directly associated to calorons, called $D\bar D$ (as opposed to $DD$). Both 
objects in such a $D\bar D$ configuration have fractional topological charged, 
but opposite in sign. Perhaps these arise when two near constituent monopoles, 
one belonging to a caloron, the other to an anti-caloron, ``annihilate".
Our analytic methods can not directly address this situation due to the 
lack of self-duality. The same holds for configurations that seem to only 
carry magnetic fields, which were already seen long ago~\cite{LaSc}. 

\begin{figure}[htb]
\vskip3.7cm
\includegraphics{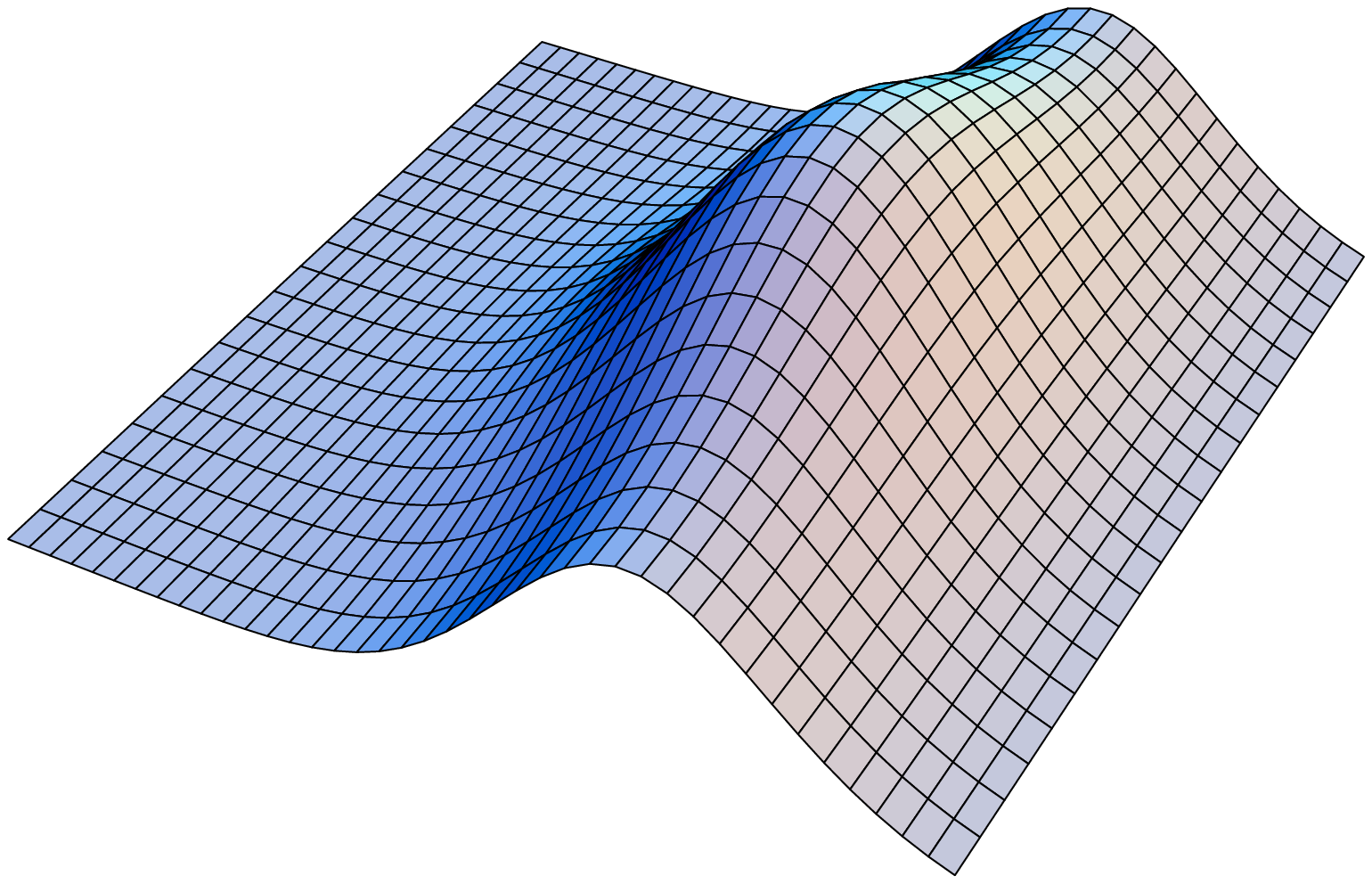}
\includegraphics{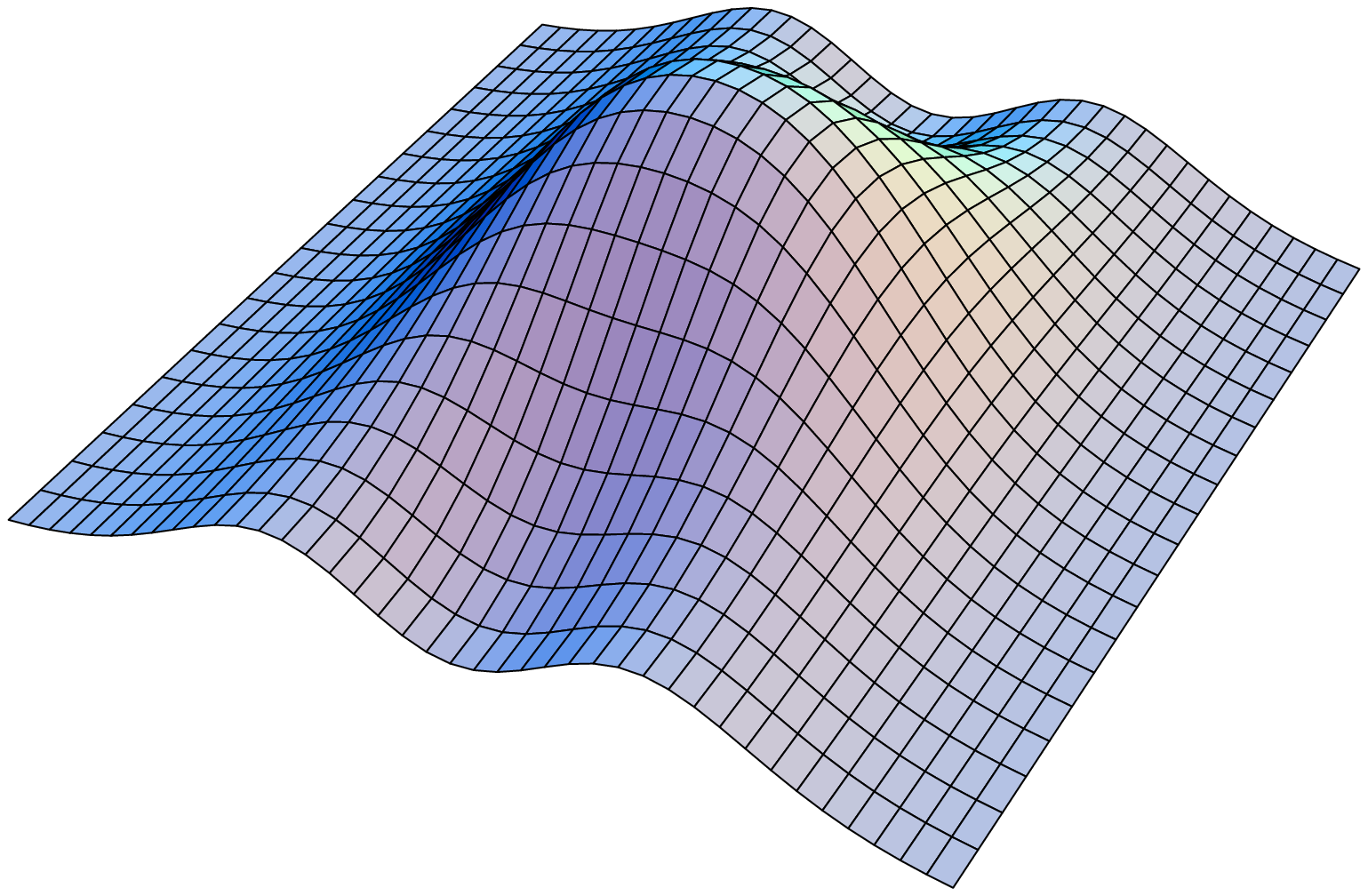}
\includegraphics{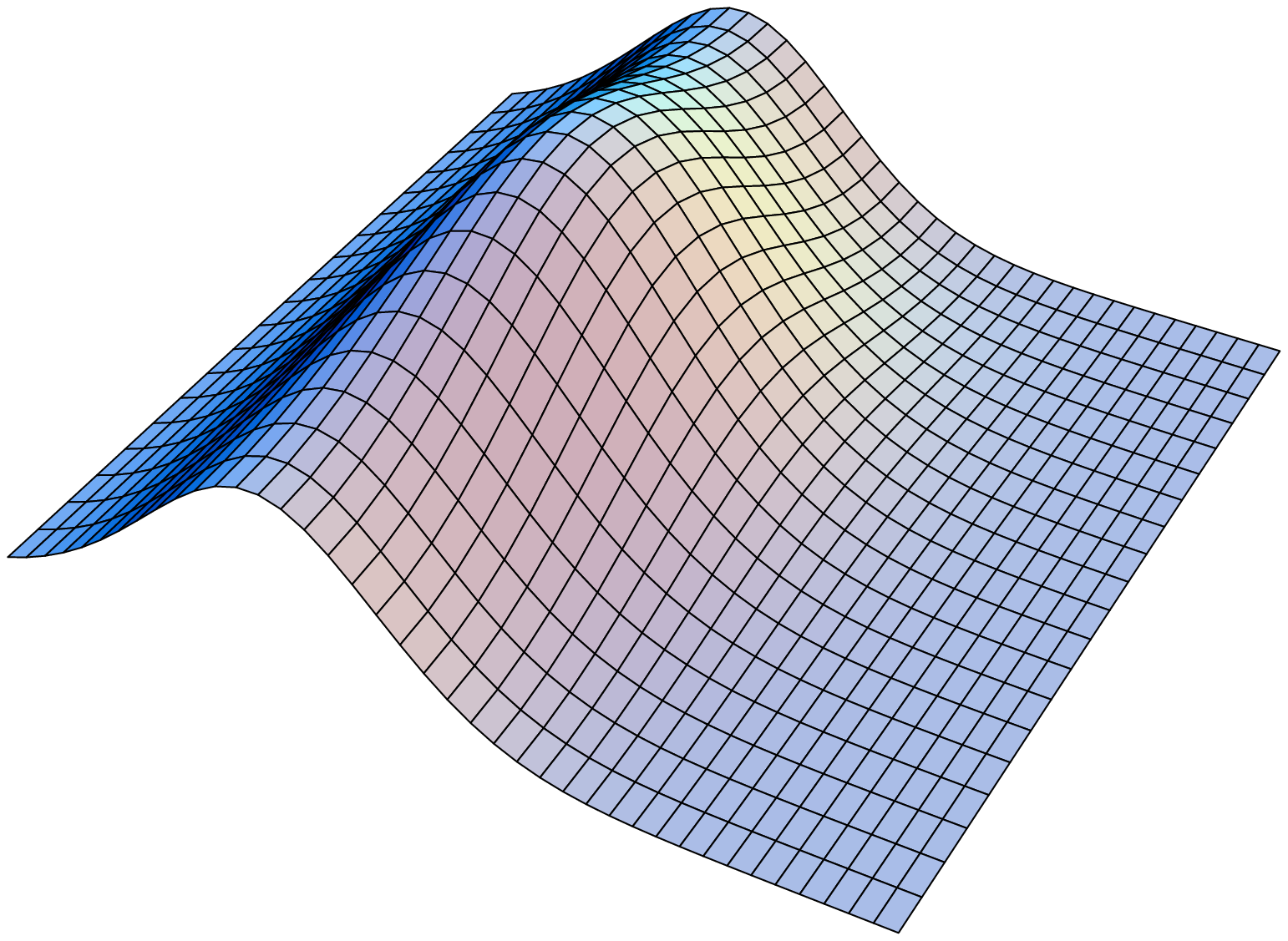}
\caption{Fermion zero-mode densities as a function of $t$ and $z$ for a 
charge 1 caloron ($\mu_2=\quarter$ and $\rho=\half$) with periodic (right) 
and anti-periodic (left) boundary conditions, compared to the action 
density (middle) (see also Fig.~5 in Ref.~\cite{Boul}). Based on
Ref.~\cite{Gattp}; produced with Ref.~\cite{WWW}).}\label{fig:zmflip}
\end{figure}

One point of criticism that applies to both methods is that the choice
of boundary conditions may force the ``dissociation" of instantons into
constituent monopoles, particularly since volumes can not yet be chosen
so large that many instantons are contained within a given configuration.
A recent study~\cite{Ilg3} has done away with the fixed boundary conditions 
that enforce the non-trivial holonomy. Nevertheless, still one finds in many 
cases that calorons ``dissociate" into constituent monopoles below the 
deconfinement transition temperature. A particularly useful tool has turned 
out to be the fermionic (near) zero-modes to detect the monopole constituents 
when they are too close together to reveal themselves from the action 
density~\cite{Ilg3}. This relies on the observation that the zero-mode 
is localized on {\em only} one of the constituent monopoles, determined 
by the boundary conditions imposed on the fermions in the time 
direction~\cite{MTCP,MTP}. For $SU(2)$ this is particularly simple,
with periodic and anti-periodic boundary conditions of the fermions making
the zero-mode switch from one constituent to the other, as is illustrated
for a close pair of constituents in Fig.~\ref{fig:zmflip}. In addition
one may use the Polyakov loop for diagnostic purposes~\cite{Ilg1,Ilg2,Ilg3}, 
for $SU(2)$ taking the values $\ein_2$ and $-\ein_2$ near the respective 
constituent locations (at these points the gauge symmetry is restored, 
providing an alternative definition for the center of a constituent monopole).

Fermion eigenfunctions with eigenvalues near zero have also been used as an 
alternative to cooling, to filter out the high frequency modes and identify 
topological lumps. For a recent lattice study, including some discussion of
calorons with non-trivial holonomy, see Ref.~\cite{Gatt} and references
therein. Using the near zero-modes as a filter, constituent monopoles have 
even been identified recently for $SU(3)$ well below the deconfining 
temperature~\cite{Gattp}, resembling Fig.~\ref{fig:zmflip} (see also Fig.~14 
of Ref.~\cite{Gatt}. 

For the exact axially symmetric multi-caloron solutions constructed in this
paper the $k$ associated fermion zero-modes (for charge $k$) will be derived
in the near future. We anticipate that one can choose a basis where each 
is localized on one of the constituent monopoles, the type of which is 
determined by the choice of fermionic boundary conditions in the time 
direction. Analyzing these zero-modes is particularly interesting in the 
light of some puzzles that were presented in a recent study~\cite{Bip} 
of the normalizable fermion zero-modes in the background of a collection 
of so-called bipoles, i.e. pairs of oppositely charged (but self-dual) 
Dirac monopoles, which are of interest in a wider context as well. This 
will be one of the many topics we have access to with our analytic tools.
But ultimately our main aim is to develop a reliable method to describe 
the long distance features of non-abelian gauge theories in terms of 
monopole constituents to understand both confinement and chiral symmetry 
breaking. The results of this paper, and in particular the recent lattice 
results, provide some encouragement in this direction.

\section*{Acknowledgements}

We thank Conor Houghton for initial collaboration on the monopole aspects
of this work and him as well as Chris Ford for extensive discussions. PvB 
also thanks Michael M\"uller-Preussker and Christof Gattringer for 
discussions concerning calorons with non-trivial holonomy on the lattice. 
Furthermore he is grateful to Leo Stodolsky and Valya Zakharov for 
hospitality at the MPI in Munich and to Poul Damgaard, Urs Heller and 
Jac Verbaarschot for inviting him to the ECT* workshop ``Non-perturbative 
Aspects of QCD'' in Trento. He thanks both institutions for their support, 
while some of the work presented in this paper was performed. FB likes to 
thank the organizers of the ``Channel Meeting on Theoretical Particle 
Physics" for a well organized and stimulating meeting as well as Dimitri 
Diakonov, Gerald Dunne, Alexander Gorsky and Peter Orland for discussions. 
The research of FB is supported by FOM.


\begin{thebibliography}{99}
\bibitem{Gattp}Christof Gattringer, private communications. The zero-mode 
densities for the case of Fig.~\ref{fig:zmflip} were produced for the purpose 
of illustrating the behavior observed by C. Gattringer and co-workers in 
$SU(3)$ lattice gauge theory.
\end{thebibliography}
\end{document}